\documentclass[a4paper, amsfonts, amssymb, amsmath, reprint, showkeys, nofootinbib, twoside, superscriptaddress, longbibliography]{revtex4-2}

\usepackage{amsmath}
\usepackage{amsfonts}
\usepackage{amssymb}
\usepackage{changes}
\usepackage[english]{babel}
\usepackage[utf8]{inputenc}
\usepackage{parskip}
\usepackage{changes}
\usepackage[pdftex, pdftitle={Article}, pdfauthor={Author}]{hyperref} 
\DeclareUnicodeCharacter{2061}{}

\newcommand{\Icp}{\ensuremath{I_\mathrm{c}^+}}
\newcommand{\Icm}{\ensuremath{I_\mathrm{c}^-}}
\newcommand{\Ir}{\ensuremath{I_\mathrm{r}}}
\newcommand{\Bz}{\ensuremath{\mu_0H_\mathrm{z}}}
\newcommand{\By}{\ensuremath{\mu_0H_\mathrm{y}}}

\newcommand{\Tcp}{\ensuremath{T_\mathrm{c}^\mathrm{P}}}
\newcommand{\Tcap}{\ensuremath{T_\mathrm{c}^\mathrm{AP}}}

\begin{document}
\title{Magnetic exchange coupled nonreciprocal devices for cryogenic memory}

\author{Josep Ingla-Ayn\'es}

    \email{jingla@mit.edu}
    \affiliation{Francis Bitter Magnet Laboratory and Plasma Science and Fusion Center,
Massachusetts Institute of Technology, Cambridge, Massachusetts 02139, USA}
\author{Lina Johnsen Kamra}
\affiliation{Department of Physics, Massachusetts Institute of Technology, Cambridge, Massachusetts 02139, USA}
\author{Franklin Dai}

    \affiliation{Newton North High School, Newton, Massachusetts 02460, USA}

\author{Yasen Hou}

    \affiliation{Francis Bitter Magnet Laboratory and Plasma Science and Fusion Center,
Massachusetts Institute of Technology, Cambridge, Massachusetts 02139, USA}

\author{Shouzhuo Yang}
    \affiliation{Francis Bitter Magnet Laboratory and Plasma Science and Fusion Center,
Massachusetts Institute of Technology, Cambridge, Massachusetts 02139, USA}
\author{Peng Chen}
    \affiliation{Francis Bitter Magnet Laboratory and Plasma Science and Fusion Center,
Massachusetts Institute of Technology, Cambridge, Massachusetts 02139, USA}
\author{Oleg A. Mukhanov}
    \affiliation{SEEQC, Inc., Elmsford, New York 10523, USA}
\author{Jagadeesh S. Moodera}
\email{moodera@mit.edu}
    \affiliation{Francis Bitter Magnet Laboratory and Plasma Science and Fusion Center,
Massachusetts Institute of Technology, Cambridge, Massachusetts 02139, USA}

\affiliation{Department of Physics, Massachusetts Institute of Technology, Cambridge, Massachusetts 02139, USA}
\date{\today} 

\begin{abstract}
As computing power demands continue to grow, superconducting electronics present an opportunity to reduce power consumption by increasing the energy efficiency of digital logic and memory. A key milestone for scaling this technology is the development of efficient superconducting memories. Such devices should be nonvolatile, scalable to high integration density and memory capacity, enable fast and low-power reading and writing operations, and be compatible with the digital logic.  
We present a versatile device platform to develop such nonvolatile memory devices consisting of an exchange-coupled ultra-thin superconductor encapsulated between two ferromagnetic insulators (FIs). The superconducting exchange coupling, which is tuneable by the relative alignment between the FI magnetizations, enables the switching of superconductivity on and off. 
{We exploit this mechanism to create a superconducting nonvolatile memory where single-cell writing is realized using heat-assisted magnetic recording, and explain how it can become a contender for state-of-the art superconducting memories.} Furthermore, below their critical temperatures, the memory elements show a marked nonreciprocity, with zero magnetic field superconducting diode efficiencies exceeding $\pm$60\%, showing the versatility of the proposed devices for superconducting computing. 
\end{abstract}


\maketitle

Cryogenic superconducting electronics has the potential to process digital data at significantly higher speed and energy efficiency than conventional semiconductor electronics. The runaway power demands of AI data centers \cite{shehabi2024} and scaling challenges for fault-tolerant quantum computing \cite{beverland2022,gidney2021} have raised further interest in its evolution. While digital superconducting logic has been steadily progressing, the absence of dense, high-capacity 
memory matching the speed and energy efficiency of superconducting logic is limiting its application \cite{alam2023}.

Despite significant research efforts in developing cryogenic memory matching superconducting single flux quantum (SFQ) circuits, the dense, high-capacity cryogenic memory remains underdeveloped. A variety of different approaches were explored \cite{alam2023}. They included memory cells based on SQUIDs \cite{nagasawa1999,semenov2019}, ferromagnet-based magnetic Josephson junctions (MJJs) \cite{baek2014,niedzielski2015,vernik2012,ryazanov2012}, cryogenic spin-torque pseudo-spin valve (PSV) \cite{ye2014} and spin-orbit magnetic elements \cite{nguyen2020}, cryoCMOS prototypes \cite{han2025}, hybrid cryoCMOS and Josephson-based circuits \cite{duzer2012,hironaka2020}, superconducting delay lines \cite{volk2023}, superconducting loops \cite{butters2021}, nano cryotrons (ntrons) \cite{buzzi2023}, Abrikosov vortex structures \cite{miyahara1987,golod2015,golod2023}, and vertically integrated Josephson junctions and PSVs or magnetic elements \cite{nevirkovets2018,nevirkovets2023}. In this context, the realization of superconducting nonvolatile memories that can be scaled to sub-$\mu$m$^2$ areas and are addressable with SFQ pulses is highly desired.

Heat-assisted magnetic recording (HAMR) enables magnetic memory writing through heating, allowing reduced magnetic-field switching of micro- and nano-magnets  \cite{kryder2008}. Recently, a HAMR-based memory has been proposed where a low Curie temperature ferromagnetic insulator (FI, EuS) is switched by this technique. Since the EuS is separated from the superconductor (SC) by a thin protective barrier, its stray fields influence the SC state \cite{pagano2017}. The nonvolatility provided by the FI makes this approach appealing for applications, but a more robust coupling between FI and SC would be desired for reliable memory operations.

\begin{figure*}[htb]
	\centering
		\includegraphics[width=0.9\textwidth]{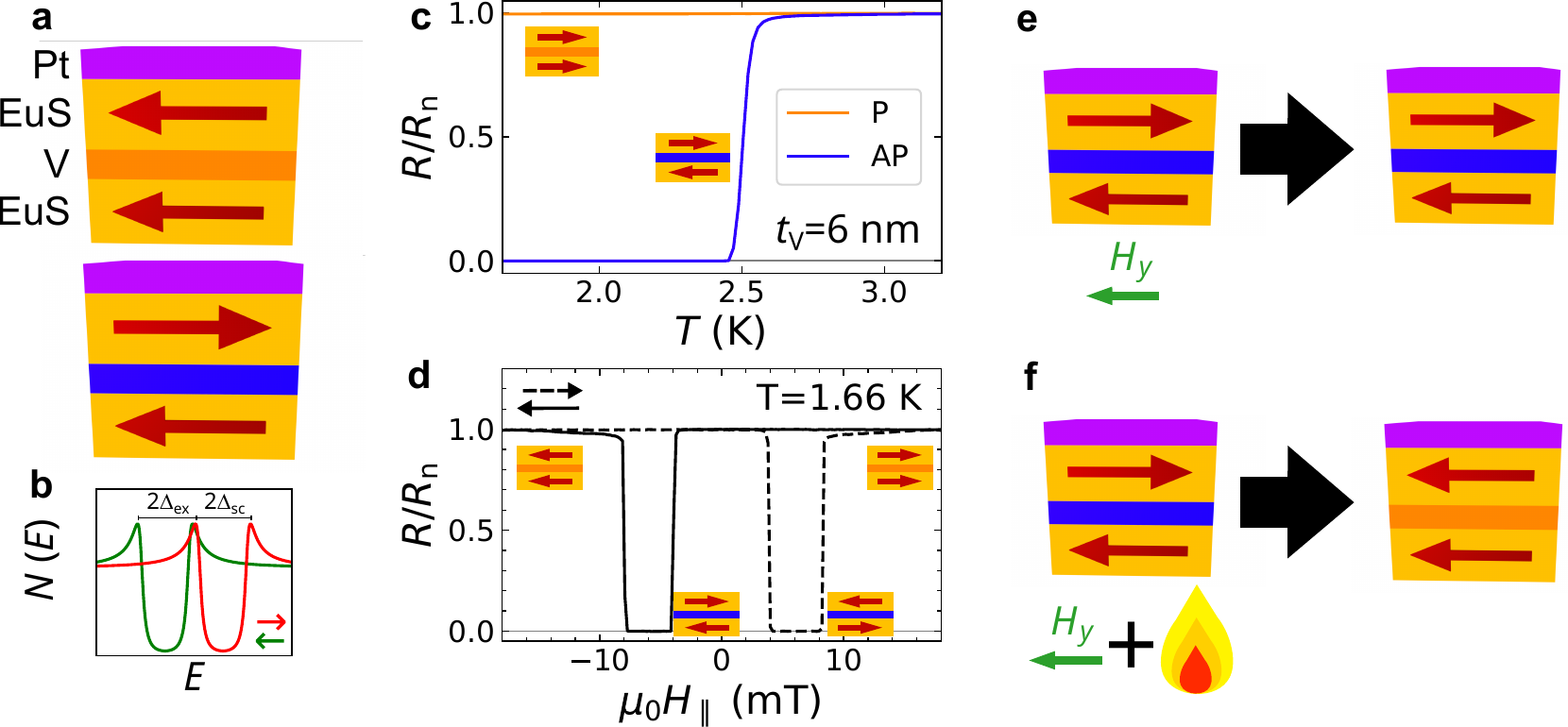}
	\caption{\textbf{Proposed superconducting memory functionality and superconducting exchange coupling in an EuS/V/EuS-based thin film multilayer.} (a) Illustration of trilayer operation. When the magnetizations are parallel (P), the exchange coupling ($\Delta_\mathrm{ex}$) —which exceeds the superconducting gap ($\Delta_\mathrm{sc}$)— disrupts the Cooper pairs in the V layer, leading to a reduction in the critical temperature ($T_c$). When the EuS magnetizations (red arrows) are anti-parallel (AP), the total $\Delta_\mathrm{ex}$ is reduced and the superconductivity in V is preserved. (b) Spin-resolved density of states in V for P magnetization alignment. (c) Experimental temperature ($T$) dependence of the resistance of a 6 nm V film in the P and AP alignment configurations illustrated by the insets. $R_\mathrm{n}$ is the normal state resistance. (d) Experimental infinite magnetoresistance where the superconductivity is switched on and off by controlling the magnetization alignment using an in-plane magnetic field ($\mu_0 H_\parallel$), establishing the foundation of the proposed nonvolatile memory. The arrows indicate the $\mu_0 H_\parallel$ sweep directions and the insets show the magnetization alignments. (e) and (f) Proposed switching mechanism. (e) A tiny magnetic field is not sufficient to switch the magnetization configuration but, when local heating increases the device temperature (f), the magnetizations become P and superconductivity is switched off.}
	\label{Figure1}
\end{figure*}

Such a robust coupling can be achieved in trilayer structures where the SC is sandwiched between two FIs (Fig.~\ref{Figure1}a). As proposed by De Gennes in 1966 \cite{degennes1966}, FI-induced magnetic exchange coupling of the SC \cite{sarma1963,bergeret2005,cai2023} enables control over the SC’s critical temperature \cite{degennes1966,hauser1969,li2013,zhu2017,dibernardo2019,matsuki2025,ojajarvi2022,bhakat2025}. When the SC is thinner than its coherence length ($\xi$), and the exchange coupling ($\Delta_\mathrm{ex}$) exceeds the pairing energy ($\Delta_\mathrm{sc}$), superconductivity is more strongly suppressed in the parallel (P) FI configuration than in the anti-parallel (AP) case (Figs.~\ref{Figure1}a-\ref{Figure1}c)  \cite{degennes1966}. Thus, at the right temperature ($T$) range, 
superconductivity can be toggled by switching the FI magnetization alignment, resulting in an infinite magnetoresistance (Fig.~\ref{Figure1}d). This phenomenon, demonstrated earlier in mm-scale thin-film multilayers \cite{hauser1969,li2013,zhu2017,dibernardo2019,matsuki2025,bhakat2025}, motivates the study of lithographically patterned devices. These may constitute new superconducting memory (SM) cells where, in contrast with \cite{pagano2017}, the entire FI-covered area undergoes a transition to the resistive state \cite{degennes1966}.  

Here, we realize a new type of SM that combines HAMR with magnetic exchange coupling. We pattern superconducting bridges of EuS/V/EuS-based multilayers and, by local heating of the EuS layers using current pulses, we switch the magnetic alignment of the EuS layers from AP to P, resulting in a transition of the proposed SMs from the SC to the resistive regime by the effect of a tiny ambient in-plane magnetic field. We demonstrate that adjacent memory cells remain superconducting and propose a path to miniaturization that should result in cooling-limited sub-ns writing times and a small area footprint in the $\mu$m$^2$ range \cite{pagano2017}, comparable to the smallest Josephson-junction based memories \cite{golod2023}. We argue that, by introducing a narrower contact for lower-current heating \cite{buzzi2023}, the proposed SMs can be fully operated by SFQ pulses. 
Furthermore, below the SC critical temperature, we demonstrate a strong nonreciprocity of our devices, which operate as highly efficient superconducting diodes (SDs) with efficiency [$\eta=(\Icp{}-|\Icm{}|)/(\Icp{}+|\Icm{}|)$, where $I_\mathrm{c}^{+(-)}$ is the forward (reverse) critical current] exceeding $\pm$60\% at zero magnetic field, with EuS magnetizations set AP. When the EuS magnetizations are P, and $T$ is below the P critical temperature (\Tcp{}), the maximal critical current [$I_\mathrm{c}^\mathrm{max}=\max⁡(\Icp{}, \Icm{})$] decreases dramatically, as dictated by the exchange coupling at the interface. The observed large zero-field SD efficiency also makes such devices suitable contenders for power delivery and logic for signal processing in SFQ circuit applications, such as data path programming and qubit control. The integration of the proposed SMs and SDs into SFQ circuits alongside Josephson junctions (JJ) will open new avenues for classical and quantum computing with higher energy efficiency, lower area overhead, and richer functionality, including digital in-memory computing \cite{ielmini2018}. 

\begin{figure}[tb]
	\centering
		\includegraphics[width=0.4\textwidth]{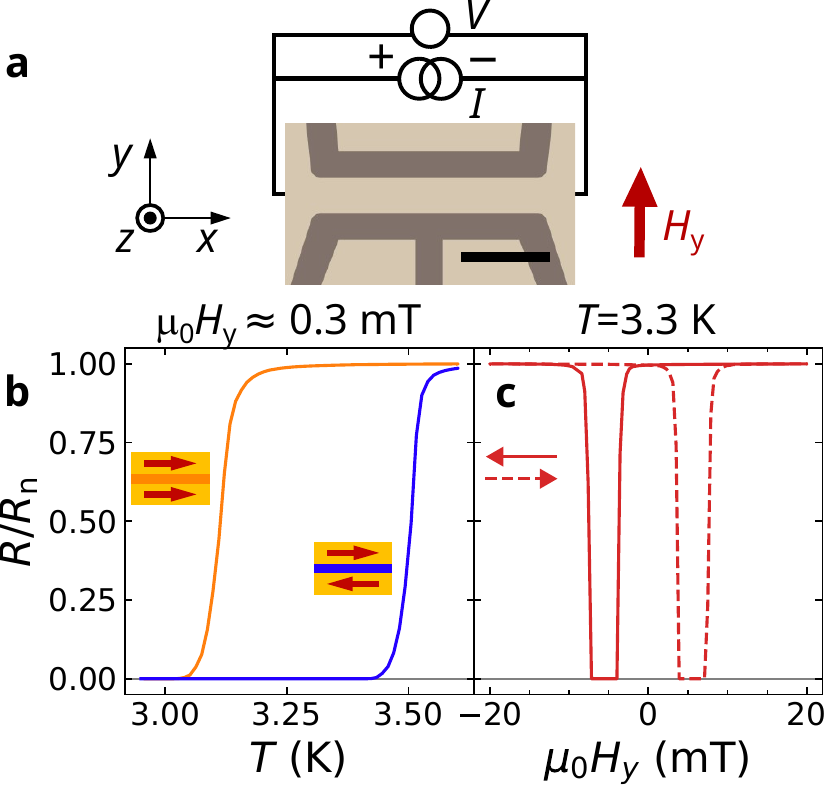}
	\caption{\textbf{Superconducting exchange coupling and infinite magnetoresistance in a patterned device.} (a) Device geometry. The darker areas are etched, and the scale bar is 20~$\mu$m. The measurement circuit (black) and the magnetic field direction (red) are also shown. (b) $T$-dependence of the device resistance in the P and AP magnetic configurations, as indicated by the red arrows in the sketches, showing a critical temperature difference $\Delta T_c\approx0.4$~K. The V thickness $t_V$ is 7~nm. (c) Infinite magnetoresistance under a magnetic field applied along $y$ (\By{}) at a temperature $T=3.3$~K. 
    The horizontal arrows indicate the \By{} sweep directions, and the magnetization alignments are shown in Fig.~\ref{Figure1}d.}
	\label{Figure2}
\end{figure}

\begin{figure}
    \centering
    \includegraphics[width=\linewidth]{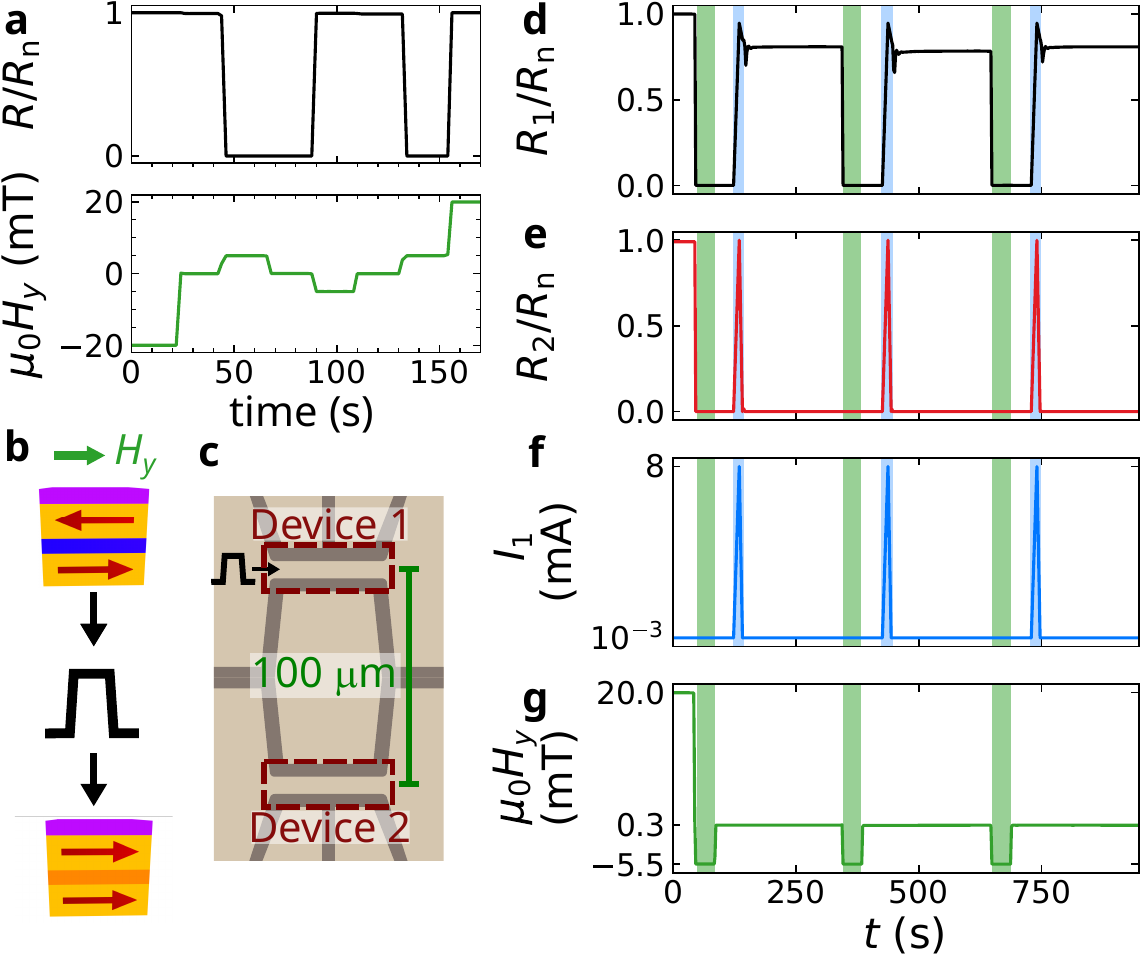}
    \caption{\textbf{Resistive switches of an individual device using current pulses.} (a) \emph{Upper panel,} nonvolatile memory functioning showing that both superconducting and resistive states are stable at zero magnetic field. \emph{Lower panel,} stepwise change of \By{} vs.~time. (b) Measurement sequence: first, the magnetizations are set AP with an external magnet and a magnetic field $\By{}\approx 0.3$~mT is applied. Next, an 8~mA current pulse is applied to switch the magnetization of the upper EuS layer, and make the 7-nm-thick V resistive. The process is repeated three times. (c) Sample geometry coloured as Fig.~\ref{Figure2}a. Devices~1 and 2, separated by 100~$\mu$m, are marked by red rectangles. The current pulse is applied only to Device~1. (d) Resistance switches between superconducting and resistive states of Device~1 ($R_1$) normalized by its $R_\mathrm{n}$. (e) Device~2 resistance ($R_2$) normalized by its $R_\mathrm{n}$. (f) Current pulses ($I_1$, blue rectangles) and (g) magnetic field pulses (green rectangles) used to induce resistive switches to Device~1 without influencing Device~2. The measurements are performed at $T=3.3$~K.}
    \label{FigureSwitching}
\end{figure}

\begin{figure*}[tb]
	\centering
		\includegraphics[width=\textwidth]{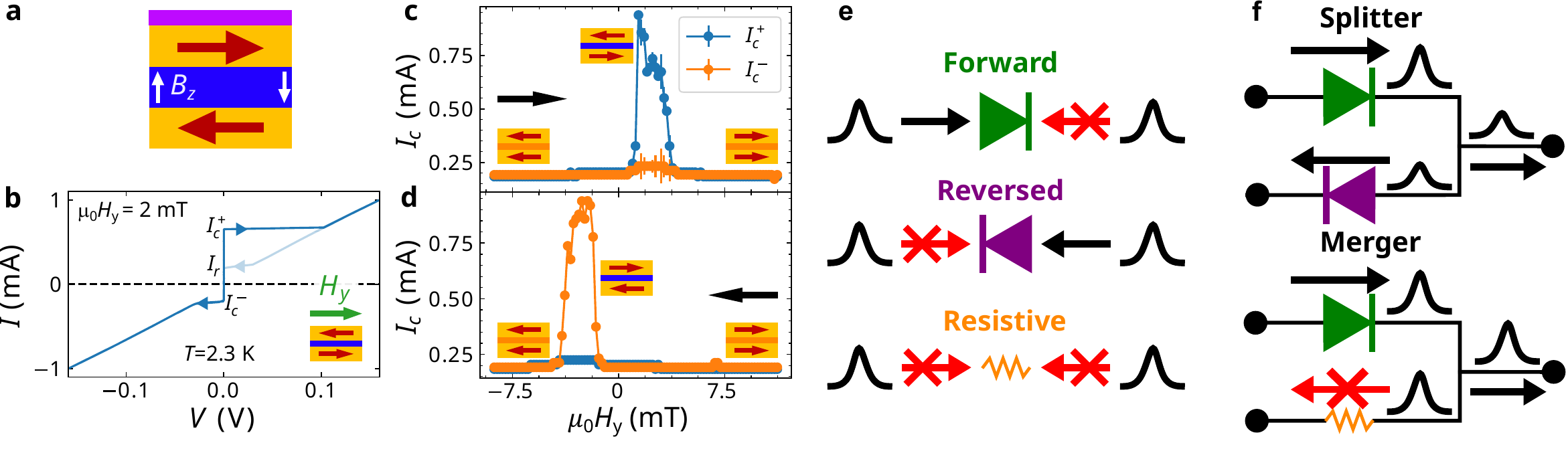}
	\caption{\textbf{Three-state superconducting diode effect at 2.3 K.} (a) Schematic illustrating the out-of-plane fringe magnetic fields ($B_z$) induced by the EuS layers in the anti-parallel alignment. The V thickness is 8~nm. (b) Measured current ($I$)-voltage ($V$) characteristics of SD1 at $\By{}=2$~mT showing the forward (\Icp{}), reverse (\Icm{}), and retrapping ($I_\mathrm{r}$) currents. (c) and (d) \Icp{} and \Icm{} as a function of \By{} at 2.3~K. The black arrows indicate the sweep direction, and the magnetization alignments are schematically shown in the insets. (e) Three-state superconducting diode (SD) with forward, reversed, and resistive (low critical current) states allows for pulse propagation towards the right, left, or blocks propagation in either direction, respectively. (f) Programmable pulse propagation using the achieved SDs. \emph{Upper panel}, a signal pulse injected through the upper arm can flow back into the lower arm when using a reversed diode to block incoming pulses, resulting in a pulse splitter. \emph{Lower panel}, when using the resistive state of the proposed SDs to block pulses, propagation in the lower arm is prevented in both forward and reverse directions, enabling a pulse merger. Both schematics showcase the enhanced flexiblility of the proposed SDs for pulse guiding schemes. The pulses in (e) and (f) represent SFQ pulses.}
	\label{Figure3}
\end{figure*}

\subsection*{Exchange control of critical temperature in superconducting memory elements}

{Despite the variety of thin-film multilayers demonstrated in the literature \cite{hauser1969,li2013,zhu2017,dibernardo2019,matsuki2025,bhakat2025}, to the best of our knowledge, the behaviour of patterned-$\mu$m scale FI/SC/FI trilayers remains unexplored.} The EuS/V/EuS devices, prepared as explained in the Methods section, are shown in Fig.~\ref{Figure2}a. The V thicknesses ($t_\mathrm{V}$, specified in each figure) and operating temperatures are chosen to optimize the desired device functionality.
The critical temperatures in the P and AP states are shown in Fig.~\ref{Figure2}b, where the magnetizations are aligned along $\pm y$. From Fig.~\ref{Figure2}b, the critical temperatures are $\Tcp{}\approx3.11$~K and $\Tcap{}\approx3.51$~K. Figure~\ref{Figure2}c shows that the SC state can be turned on and off by switching the magnetic alignment. Figure~\ref{Figure2}c shows the \By{}-dependent infinite magnetoresistance at $T=3.3$~K, a major requirement for SM applications. 
Note that, even though results in exfoliated FI/SC/FI devices have been obtained in Ref.~\cite{yun2023}, no evidence for $\Tcp{}\neq\Tcap{}$ is presented.

\subsection*{Current-assisted switching of individual superconducting memory elements}

An important application of FI/SC/FI trilayers where superconductivity can be switched on and off in a nonvolatile way is superconducting memory. The perfect nonvolatile memory functioning of one of our EuS/V/EuS-based devices is shown in Fig.~\ref{FigureSwitching}a where the resistance of Device~1 ($R_1$, see Fig.\ref{FigureSwitching}c, Device~2 shows practically identical performance), which is plotted vs.~time, goes from zero to a finite value ($R_\mathrm{n}\approx 215\,\Omega$) when switching from AP to P state using \By{} (Fig.~\ref{FigureSwitching}a, lower panel). It is worth noting that, once \By{} has changed the state between P and AP, the device maintains its resistance value when the field is brought back to zero, giving rise to two remanent states with infinitely different resistances \cite{li2013,bhakat2025}.

A crucial requirement for the integration of such perfect SMs into useful circuits is the switching of individual devices. For this purpose, we apply electrical current pulses (Figs.~\ref{FigureSwitching}b and \ref{FigureSwitching}c). As shown in Fig.~\ref{FigureSwitching}d, at $T=3.3$~K, which is between \Tcp{} and \Tcap{}, the application of a current pulse to Device~1 with 8~mA current (corresponding to a current density of $1.4\times10^7$~A/cm$^2$) and 1~s duration results in an infinite change in the device resistance, from the superconducting to the resistive state. Since this pulse keeps the adjacent Device~2 unchanged (Fig.~\ref{FigureSwitching}e), Figure~\ref{FigureSwitching} demonstrates the selective switching of a single device. The switching is mostly mediated by heating caused by the current pulse, which reduces the coercivity of Device~1 more than that of Device~2 \cite{pagano2017}. As a result, a tiny (in this case $\By{}\approx0.3$~mT) magnetic field is sufficient to align the EuS magnetizations P. Figure~\ref{FigureSwitching}d also shows small resistance oscillations immediately after the current pulses. We attribute these oscillations to small temperature fluctuations in the measurement setup, which are caused by the high thermal conductance of the sapphire substrates. More details on the switching can be found in the Supplementary Information Section~S1.

Figure~\ref{FigureSwitching} represents a crucial step towards the realization of superconducting memories with readout possible at arbitrarily low bias current. {We compare this result with a recently proposed SC/ferromagnet-based memory device where spin-orbit induced magnetization switching has been realized \cite{cheng2025} to reverse an SD and find that both the HAMR and reading proposed here require significantly lower currents.} Although our heat-assisted writing only allows for switching individual memory cells towards the P alignment dictated by the magnetic field, implying that the magnetic field must be reversed to set the memories back to their AP state, the required magnetic fields are small and may be created on-chip by micromagnets. Thus, the proposed SMs may be used in various settings. For example, memory applications that require frequent reading and rare writing, such as read-only memory (ROM) or programmable auxiliary elements to simplify the SFQ circuits \cite{mukhanov1987}, or for digital in-memory computing. 

We would like to remark that the effect can be greatly optimized by replacing the sapphire substrates with SiO$_2$. Due to its high thermal conductivity, sapphire requires additional heating power. In contrast, SiO$_2$ has a significantly lower thermal conductivity, which reduces the required heat input and enables faster, more localized heating, as being exploited by current volatile SM prototypes \cite{buzzi2023}. 

{Compatibility with superconducting logic circuits implies that SFQ pulses, when applied to the correct terminal, should be used for writing and reading. Our planar devices can be easily optimized using a narrower contact with a low critical current for memory writing. In this case, a train of SFQ pulses (equivalent of applying $\sim$300~$\mu$A current) to the write terminal would increase the memory cell temperature and enable switching. A single SFQ pulse applied to the main terminal would perform reading of the memory state using very low energy.}

\subsection*{SD effect with magnetic exchange coupling}

Figure~\ref{Figure3}a shows a schematic configuration of the out-of-plane fringe fields in our EuS/V/EuS-based multilayer devices when the magnetization alignment is AP. In this case, the out-of-plane fringe fields from the upper and lower EuS layers add up at the edges. For this reason, an efficient SD effect beyond the state of the art in FI/SC systems \cite{hou2023, suri2022, ingla2025} is expected. Data in Figs.~\ref{Figure3}b-\ref{Figure3}d is taken with the superconducting diode at a temperature of 2.3~K, that is below \Tcp{} (see Fig.~\ref{Figure2}b) where $\Delta_\mathrm{sc}$ dominates $\Delta_\mathrm{ex}$. Figure~\ref{Figure3}b shows the observed SD effect in a typical current ($I$)-voltage ($V$) characteristic curve where the arrows indicate the sweep direction and forward and reverse critical and retrapping currents (\Icp{}, \Icm{}, and \Ir{}, respectively) are defined. The EuS magnetization orientation and \By{} are shown in the inset. Figures~\ref{Figure3}c and \ref{Figure3}d show the critical currents ($I_c$) as a function of \By{}, that is swept in the direction indicated by the black arrows. As shown here, when the EuS magnetizations are AP, either \Icp{} or \Icm{} becomes greater than 0.5~mA, depending on the magnetic configuration. As shown by the inset sketches, the AP configurations are inverted in Fig.~\ref{Figure3}c and \ref{Figure3}d, resulting in reversed stray fields and diode efficiencies with opposite sign. As a result, the maximum SD efficiency in Fig.~\ref{Figure3}c is $\eta\approx57$\% and, in Fig.~\ref{Figure3}d is $\eta\approx-58$\%, state-of-the-art values in the SD literature \cite{hou2023}. These results contrast with the P configuration, where both critical currents are below 0.25~mA, attributable to the weakened superconducting coupling (see Supplementary Information Section~S2 for
results at lower $T$). In this case, $\eta$ is practically zero and, for currents between 0.25 and 0.6~mA, our device is in the resistive state, demonstrating the magnetic alignment dependence of our multifunctional device. The realization of three-state programmable SDs (Figs.~\ref{Figure3}c and \ref{Figure3}d) capable of routing data selectively to the left, right, or blocking any flow (Fig.~\ref{Figure3}e) would open new possibilities, for superconducting programmable logic circuits 
(Fig.~\ref{Figure3}f). 

Table~\ref{Table1} shows the zero-field SD efficiencies in the forward ($\eta_+>0$) and reverse ($\eta_-<0$) configurations of four different SDs (SD1 to SD4) patterned on two films grown in different batches, highlighting the reproducibility of our work, and their suitability for power delivery applications \cite{nadeem2023,castellani2025,ingla2025}. In addition, we obtained $\eta\approx\pm80$\% in an SD fabricated using 7~nm V, demonstrating that the $\eta\approx\pm60$\% range is not the ultimate limit of the presented devices (see Supplementary Information Sections~S2-S4 for the complete characterization of all the SDs). Considering that the state-of-the-art zero-field SD performance for scalable SDs based on FI/SC bilayers 
has recently reached values up to 54\% \cite{li2025} using perpendicularly magnetized Co and edge asymmetry, and the best scalable $\eta$ is $\approx70$\% in Josephson junctions \cite{golod2022}, our scalable devices with efficiencies reaching 80\% represent a significant step forward for SD applications. The variability between samples is enhanced by the asymmetry between edges \cite{hou2023, 2vodolazov2005}, resulting in maximum $\eta$ values occurring at $\mu_0 H \neq 0$. In Supplementary Information Section~S5 we compare the effects of an asymmetry in the EuS edge magnetizations with those of an asymmetric surface barrier in the V layer.
These solvable limitations can easily be addressed with better control of the edge-definition process and by choosing a less sensitive SC layer.
\begin{table}[htb]
\caption{Reproducibility of highly efficient SD effect in EuS/V/EuS-based multilayers with 8~nm V at \( T = 1.7\,\mathrm{K} \) and \( \mu_0 H = 0 \).}
\centering
\begin{tabular}{lcc}
\hline
Sample & \( \eta_+\,(\%)\) & \( \eta_-\,(\%) \) \\
\hline
SD1 & \( 37 \pm 3 \) & \( -47 \pm 2 \) \\
SD2 & \( 53 \pm 3 \) & \( -59 \pm 1 \) \\
SD3 & \( 65 \pm 2 \) & \( -54 \pm 3 \) \\
SD4 & \( 66.5 \pm 0.9 \) & \( -61 \pm 2 \) \\
\hline
\end{tabular}\label{Table1}
\end{table}

To conclude, we have investigated exchange-coupled EuS/V/EuS thin-film devices where the critical temperature of the superconducting V layer depends on the EuS relative magnetization alignment.  
Such a platform has allowed us to realize (i) superconducting nonvolatile memories with heat-assisted magnetization switching, (ii) scalable SDs with record efficiencies exceeding 60\% at zero magnetic field, (iii) a new SD device concept that has three remnant states: forward, reverse, and low $I_c$.   
These results represent a crucial step for cryogenic memory and logic applications. 
The integration of the proposed SMs and SDs into SFQ circuits alongside Josephson junctions (JJ) will open new avenues for classical and quantum computing with higher energy efficiency, lower area overhead, and richer functionality, including data-path programming and qubit control in SFQ systems, and digital in-memory computing applications \cite{ielmini2018}. Furthermore, new exciting possibilities arise when combining the existing devices with our earlier observation of Majorana bound states in V/Au/EuS \cite{manna2020}. The rich physics and potential present in these scalable systems may enable the on-chip integration of the proposed logic devices with topological qubits, providing a new platform for quantum computing.

\section*{Methods}\label{SecMethods}
{The EuS/V/EuS-based films with V thicknesses ranging from 6 to 8~nm (specified in each figure), 5-nm-thick EuS layers, and covered by a 3-nm-thick Pt capping layer, were grown on sapphire substrates by e-beam evaporation in a molecular beam epitaxy system  with a base pressure below $4\times10^{-10}$~Torr. The ultra-thin film layer stacks were patterned to a width of 8~$\mu$m using e-beam lithography and ion milling. SD1 and SD2 are on the same substrate with 8-nm-thick V, SD3 and SD4 are in another film with the same V thickness. Devices~1 and 2, shown in Fig.~\ref{FigureSwitching} as well as SD5, which corresponds to Device~1 in an earlier cooldown (see Supplementary Information Section~S4 for its SD performance showing $\eta$ up to $\pm80$\%), are on a third substrate with 7-nm-thick V.} 

{The samples were cooled in a DynaCool physical property measurement system (PPMS) and measured using a rotator to apply magnetic fields along the $y$ and $z$ directions. The superconducting coils trap magnetic flux, resulting in uncertainty in the actual applied magnetic field. In Fig.~\ref{FigureSwitching}, to reduce the presence of trapped flux in the superconducting magnet, we followed the field oscillation procedure of the PPMS. For the SD measurements, we extracted the magnetic field offset using the $R$ vs.~\By{} traces, such as the ones shown in Fig.~\ref{Figure2}c. These measurements contain two traces, $R_\mathrm{inc}$ and $R_\mathrm{dec}$, corresponding to increasing and decreasing \By{} sweep directions, respectively. Since these curves are expected to follow $R_\mathrm{inc}(\By{})-R_\mathrm{dec}(-\By{})=0$, we fit for an offset in \By{} that minimized the difference between both traces. This approach allowed us to obtain an accurate estimate for the offset magnetic field right before the SD measurements, as shown in Supplementary Information Sections~S2A and S3A, enabling the determination of $\eta(\Bz{}=0)$.}

{All the data except for the $I$-$V$ traces were measured using the resistivity option from Quantum Design. The $I$-$V$ traces required to extract the critical currents were measured using a Keithley 2450 in a four-wire current source and voltage measurement mode.}

\section{Acknowledgements}
We acknowledge M. Mondol for technical assistance. This work was supported by Air Force Office of Sponsored Research (grant no. FA9550-23-1-0004 DEF), National Science Foundation (grant no. NSF-DMR 2218550), Army Research Office (grant nos. W911NF-20-2-0061, DURIP W911NF-20-1-0074), and by the MIT’s Lincoln Laboratory Advanced Concepts Committee (ACC 857). This work was carried out in part through the use of MIT.nano’s facilities.

\bibliography{bibliography}

\begin{thebibliography}{49}%
\makeatletter
\providecommand \@ifxundefined [1]{%
 \@ifx{#1\undefined}
}%
\providecommand \@ifnum [1]{%
 \ifnum #1\expandafter \@firstoftwo
 \else \expandafter \@secondoftwo
 \fi
}%
\providecommand \@ifx [1]{%
 \ifx #1\expandafter \@firstoftwo
 \else \expandafter \@secondoftwo
 \fi
}%
\providecommand \natexlab [1]{#1}%
\providecommand \enquote  [1]{``#1''}%
\providecommand \bibnamefont  [1]{#1}%
\providecommand \bibfnamefont [1]{#1}%
\providecommand \citenamefont [1]{#1}%
\providecommand \href@noop [0]{\@secondoftwo}%
\providecommand \href [0]{\begingroup \@sanitize@url \@href}%
\providecommand \@href[1]{\@@startlink{#1}\@@href}%
\providecommand \@@href[1]{\endgroup#1\@@endlink}%
\providecommand \@sanitize@url [0]{\catcode `\\12\catcode `\$12\catcode `\&12\catcode `\#12\catcode `\^12\catcode `\_12\catcode `\%12\relax}%
\providecommand \@@startlink[1]{}%
\providecommand \@@endlink[0]{}%
\providecommand \url  [0]{\begingroup\@sanitize@url \@url }%
\providecommand \@url [1]{\endgroup\@href {#1}{\urlprefix }}%
\providecommand \urlprefix  [0]{URL }%
\providecommand \Eprint [0]{\href }%
\providecommand \doibase [0]{https://doi.org/}%
\providecommand \selectlanguage [0]{\@gobble}%
\providecommand \bibinfo  [0]{\@secondoftwo}%
\providecommand \bibfield  [0]{\@secondoftwo}%
\providecommand \translation [1]{[#1]}%
\providecommand \BibitemOpen [0]{}%
\providecommand \bibitemStop [0]{}%
\providecommand \bibitemNoStop [0]{.\EOS\space}%
\providecommand \EOS [0]{\spacefactor3000\relax}%
\providecommand \BibitemShut  [1]{\csname bibitem#1\endcsname}%
\let\auto@bib@innerbib\@empty
\bibitem [{\citenamefont {Shehabi}\ \emph {et~al.}(2024)\citenamefont {Shehabi}, \citenamefont {Hubbard}, \citenamefont {Newkirk}, \citenamefont {Lei}, \citenamefont {Siddik}, \citenamefont {Holecek}, \citenamefont {Koomey}, \citenamefont {Masanet}, \citenamefont {Sartor} \emph {et~al.}}]{shehabi2024}%
  \BibitemOpen
  \bibfield  {author} {\bibinfo {author} {\bibfnamefont {A.}~\bibnamefont {Shehabi}}, \bibinfo {author} {\bibfnamefont {A.}~\bibnamefont {Hubbard}}, \bibinfo {author} {\bibfnamefont {A.}~\bibnamefont {Newkirk}}, \bibinfo {author} {\bibfnamefont {N.}~\bibnamefont {Lei}}, \bibinfo {author} {\bibfnamefont {M.~A.~B.}\ \bibnamefont {Siddik}}, \bibinfo {author} {\bibfnamefont {B.}~\bibnamefont {Holecek}}, \bibinfo {author} {\bibfnamefont {J.}~\bibnamefont {Koomey}}, \bibinfo {author} {\bibfnamefont {E.}~\bibnamefont {Masanet}}, \bibinfo {author} {\bibfnamefont {D.}~\bibnamefont {Sartor}}, \emph {et~al.},\ }\href {https://eta-publications.lbl.gov/publications/2024-lbnl-data-center-energy-usage-report} {\emph {\bibinfo {title} {2024 united states data center energy usage report}}},\ \bibinfo {type} {Tech. Rep.}\ (\bibinfo  {institution} {Lawrence Berkeley National Laboratory},\ \bibinfo {year} {2024})\BibitemShut {NoStop}%
\bibitem [{\citenamefont {Beverland}\ \emph {et~al.}(2022)\citenamefont {Beverland}, \citenamefont {Murali}, \citenamefont {Troyer}, \citenamefont {Svore}, \citenamefont {Hoefler}, \citenamefont {Kliuchnikov}, \citenamefont {Low}, \citenamefont {Soeken}, \citenamefont {Sundaram},\ and\ \citenamefont {Vaschillo}}]{beverland2022}%
  \BibitemOpen
  \bibfield  {author} {\bibinfo {author} {\bibfnamefont {M.~E.}\ \bibnamefont {Beverland}}, \bibinfo {author} {\bibfnamefont {P.}~\bibnamefont {Murali}}, \bibinfo {author} {\bibfnamefont {M.}~\bibnamefont {Troyer}}, \bibinfo {author} {\bibfnamefont {K.~M.}\ \bibnamefont {Svore}}, \bibinfo {author} {\bibfnamefont {T.}~\bibnamefont {Hoefler}}, \bibinfo {author} {\bibfnamefont {V.}~\bibnamefont {Kliuchnikov}}, \bibinfo {author} {\bibfnamefont {G.~H.}\ \bibnamefont {Low}}, \bibinfo {author} {\bibfnamefont {M.}~\bibnamefont {Soeken}}, \bibinfo {author} {\bibfnamefont {A.}~\bibnamefont {Sundaram}},\ and\ \bibinfo {author} {\bibfnamefont {A.}~\bibnamefont {Vaschillo}},\ }\bibfield  {title} {\bibinfo {title} {Assessing requirements to scale to practical quantum advantage},\ }\href {https://arxiv.org/abs/2211.07629} {\bibfield  {journal} {\bibinfo  {journal} {arXiv preprint arXiv:2211.07629}\ } (\bibinfo {year} {2022})}\BibitemShut {NoStop}%
\bibitem [{\citenamefont {Gidney}\ and\ \citenamefont {Eker{\aa}}(2021)}]{gidney2021}%
  \BibitemOpen
  \bibfield  {author} {\bibinfo {author} {\bibfnamefont {C.}~\bibnamefont {Gidney}}\ and\ \bibinfo {author} {\bibfnamefont {M.}~\bibnamefont {Eker{\aa}}},\ }\bibfield  {title} {\bibinfo {title} {How to factor 2048 bit rsa integers in 8 hours using 20 million noisy qubits},\ }\href {https://doi.org/10.22331/q-2021-04-15-433} {\bibfield  {journal} {\bibinfo  {journal} {Quantum}\ }\textbf {\bibinfo {volume} {5}},\ \bibinfo {pages} {433} (\bibinfo {year} {2021})}\BibitemShut {NoStop}%
\bibitem [{\citenamefont {Alam}\ \emph {et~al.}(2023)\citenamefont {Alam}, \citenamefont {Hossain}, \citenamefont {Srinivasa},\ and\ \citenamefont {Aziz}}]{alam2023}%
  \BibitemOpen
  \bibfield  {author} {\bibinfo {author} {\bibfnamefont {S.}~\bibnamefont {Alam}}, \bibinfo {author} {\bibfnamefont {M.~S.}\ \bibnamefont {Hossain}}, \bibinfo {author} {\bibfnamefont {S.~R.}\ \bibnamefont {Srinivasa}},\ and\ \bibinfo {author} {\bibfnamefont {A.}~\bibnamefont {Aziz}},\ }\bibfield  {title} {\bibinfo {title} {Cryogenic memory technologies},\ }\href {https://www.nature.com/articles/s41928-023-00930-2} {\bibfield  {journal} {\bibinfo  {journal} {Nature Electronics}\ }\textbf {\bibinfo {volume} {6}},\ \bibinfo {pages} {185} (\bibinfo {year} {2023})}\BibitemShut {NoStop}%
\bibitem [{\citenamefont {Nagasawa}\ \emph {et~al.}(1999)\citenamefont {Nagasawa}, \citenamefont {Numata}, \citenamefont {Hashimoto},\ and\ \citenamefont {Tahara}}]{nagasawa1999}%
  \BibitemOpen
  \bibfield  {author} {\bibinfo {author} {\bibfnamefont {S.}~\bibnamefont {Nagasawa}}, \bibinfo {author} {\bibfnamefont {H.}~\bibnamefont {Numata}}, \bibinfo {author} {\bibfnamefont {Y.}~\bibnamefont {Hashimoto}},\ and\ \bibinfo {author} {\bibfnamefont {S.}~\bibnamefont {Tahara}},\ }\bibfield  {title} {\bibinfo {title} {High-frequency clock operation of josephson 256-word/spl times/16-bit rams},\ }\href {https://ieeexplore.ieee.org/document/783834} {\bibfield  {journal} {\bibinfo  {journal} {IEEE transactions on applied superconductivity}\ }\textbf {\bibinfo {volume} {9}},\ \bibinfo {pages} {3708} (\bibinfo {year} {1999})}\BibitemShut {NoStop}%
\bibitem [{\citenamefont {Semenov}\ \emph {et~al.}(2019)\citenamefont {Semenov}, \citenamefont {Polyakov},\ and\ \citenamefont {Tolpygo}}]{semenov2019}%
  \BibitemOpen
  \bibfield  {author} {\bibinfo {author} {\bibfnamefont {V.~K.}\ \bibnamefont {Semenov}}, \bibinfo {author} {\bibfnamefont {Y.~A.}\ \bibnamefont {Polyakov}},\ and\ \bibinfo {author} {\bibfnamefont {S.~K.}\ \bibnamefont {Tolpygo}},\ }\bibfield  {title} {\bibinfo {title} {Very large scale integration of josephson-junction-based superconductor random access memories},\ }\href {https://ieeexplore.ieee.org/abstract/document/8667388} {\bibfield  {journal} {\bibinfo  {journal} {IEEE Transactions on Applied Superconductivity}\ }\textbf {\bibinfo {volume} {29}},\ \bibinfo {pages} {1} (\bibinfo {year} {2019})}\BibitemShut {NoStop}%
\bibitem [{\citenamefont {Baek}\ \emph {et~al.}(2014)\citenamefont {Baek}, \citenamefont {Rippard}, \citenamefont {Benz}, \citenamefont {Russek},\ and\ \citenamefont {Dresselhaus}}]{baek2014}%
  \BibitemOpen
  \bibfield  {author} {\bibinfo {author} {\bibfnamefont {B.}~\bibnamefont {Baek}}, \bibinfo {author} {\bibfnamefont {W.~H.}\ \bibnamefont {Rippard}}, \bibinfo {author} {\bibfnamefont {S.~P.}\ \bibnamefont {Benz}}, \bibinfo {author} {\bibfnamefont {S.~E.}\ \bibnamefont {Russek}},\ and\ \bibinfo {author} {\bibfnamefont {P.~D.}\ \bibnamefont {Dresselhaus}},\ }\bibfield  {title} {\bibinfo {title} {Hybrid superconducting-magnetic memory device using competing order parameters},\ }\href {https://www.nature.com/articles/ncomms4888} {\bibfield  {journal} {\bibinfo  {journal} {Nature Communications}\ }\textbf {\bibinfo {volume} {5}},\ \bibinfo {pages} {3888} (\bibinfo {year} {2014})}\BibitemShut {NoStop}%
\bibitem [{\citenamefont {Niedzielski}\ \emph {et~al.}(2015)\citenamefont {Niedzielski}, \citenamefont {Gingrich}, \citenamefont {Loloee}, \citenamefont {Pratt},\ and\ \citenamefont {Birge}}]{niedzielski2015}%
  \BibitemOpen
  \bibfield  {author} {\bibinfo {author} {\bibfnamefont {B.~M.}\ \bibnamefont {Niedzielski}}, \bibinfo {author} {\bibfnamefont {E.}~\bibnamefont {Gingrich}}, \bibinfo {author} {\bibfnamefont {R.}~\bibnamefont {Loloee}}, \bibinfo {author} {\bibfnamefont {W.}~\bibnamefont {Pratt}},\ and\ \bibinfo {author} {\bibfnamefont {N.~O.}\ \bibnamefont {Birge}},\ }\bibfield  {title} {\bibinfo {title} {S/f/s josephson junctions with single-domain ferromagnets for memory applications},\ }\href {https://iopscience.iop.org/article/10.1088/0953-2048/28/8/085012} {\bibfield  {journal} {\bibinfo  {journal} {Superconductor Science and Technology}\ }\textbf {\bibinfo {volume} {28}},\ \bibinfo {pages} {085012} (\bibinfo {year} {2015})}\BibitemShut {NoStop}%
\bibitem [{\citenamefont {Vernik}\ \emph {et~al.}(2012)\citenamefont {Vernik}, \citenamefont {Bol'ginov}, \citenamefont {Bakurskiy}, \citenamefont {Golubov}, \citenamefont {Kupriyanov}, \citenamefont {Ryazanov},\ and\ \citenamefont {Mukhanov}}]{vernik2012}%
  \BibitemOpen
  \bibfield  {author} {\bibinfo {author} {\bibfnamefont {I.~V.}\ \bibnamefont {Vernik}}, \bibinfo {author} {\bibfnamefont {V.~V.}\ \bibnamefont {Bol'ginov}}, \bibinfo {author} {\bibfnamefont {S.~V.}\ \bibnamefont {Bakurskiy}}, \bibinfo {author} {\bibfnamefont {A.~A.}\ \bibnamefont {Golubov}}, \bibinfo {author} {\bibfnamefont {M.~Y.}\ \bibnamefont {Kupriyanov}}, \bibinfo {author} {\bibfnamefont {V.~V.}\ \bibnamefont {Ryazanov}},\ and\ \bibinfo {author} {\bibfnamefont {O.~A.}\ \bibnamefont {Mukhanov}},\ }\bibfield  {title} {\bibinfo {title} {Magnetic josephson junctions with superconducting interlayer for cryogenic memory},\ }\href {https://ieeexplore.ieee.org/document/6377273} {\bibfield  {journal} {\bibinfo  {journal} {IEEE transactions on applied superconductivity}\ }\textbf {\bibinfo {volume} {23}},\ \bibinfo {pages} {1701208} (\bibinfo {year} {2012})}\BibitemShut {NoStop}%
\bibitem [{\citenamefont {Ryazanov}\ \emph {et~al.}(2012)\citenamefont {Ryazanov}, \citenamefont {Bol’ginov}, \citenamefont {Sobanin}, \citenamefont {Vernik}, \citenamefont {Tolpygo}, \citenamefont {Kadin},\ and\ \citenamefont {Mukhanov}}]{ryazanov2012}%
  \BibitemOpen
  \bibfield  {author} {\bibinfo {author} {\bibfnamefont {V.~V.}\ \bibnamefont {Ryazanov}}, \bibinfo {author} {\bibfnamefont {V.~V.}\ \bibnamefont {Bol’ginov}}, \bibinfo {author} {\bibfnamefont {D.~S.}\ \bibnamefont {Sobanin}}, \bibinfo {author} {\bibfnamefont {I.~V.}\ \bibnamefont {Vernik}}, \bibinfo {author} {\bibfnamefont {S.~K.}\ \bibnamefont {Tolpygo}}, \bibinfo {author} {\bibfnamefont {A.~M.}\ \bibnamefont {Kadin}},\ and\ \bibinfo {author} {\bibfnamefont {O.~A.}\ \bibnamefont {Mukhanov}},\ }\bibfield  {title} {\bibinfo {title} {Magnetic josephson junction technology for digital and memory applications},\ }\href {https://www.sciencedirect.com/science/article/pii/S1875389212020639} {\bibfield  {journal} {\bibinfo  {journal} {Physics Procedia}\ }\textbf {\bibinfo {volume} {36}},\ \bibinfo {pages} {35} (\bibinfo {year} {2012})}\BibitemShut {NoStop}%
\bibitem [{\citenamefont {Ye}\ \emph {et~al.}(2014)\citenamefont {Ye}, \citenamefont {Gopman}, \citenamefont {Rehm}, \citenamefont {Backes}, \citenamefont {Wolf}, \citenamefont {Ohki}, \citenamefont {Kirichenko}, \citenamefont {Vernik}, \citenamefont {Mukhanov},\ and\ \citenamefont {Kent}}]{ye2014}%
  \BibitemOpen
  \bibfield  {author} {\bibinfo {author} {\bibfnamefont {L.}~\bibnamefont {Ye}}, \bibinfo {author} {\bibfnamefont {D.}~\bibnamefont {Gopman}}, \bibinfo {author} {\bibfnamefont {L.}~\bibnamefont {Rehm}}, \bibinfo {author} {\bibfnamefont {D.}~\bibnamefont {Backes}}, \bibinfo {author} {\bibfnamefont {G.}~\bibnamefont {Wolf}}, \bibinfo {author} {\bibfnamefont {T.}~\bibnamefont {Ohki}}, \bibinfo {author} {\bibfnamefont {A.}~\bibnamefont {Kirichenko}}, \bibinfo {author} {\bibfnamefont {I.}~\bibnamefont {Vernik}}, \bibinfo {author} {\bibfnamefont {O.}~\bibnamefont {Mukhanov}},\ and\ \bibinfo {author} {\bibfnamefont {A.}~\bibnamefont {Kent}},\ }\bibfield  {title} {\bibinfo {title} {Spin-transfer switching of orthogonal spin-valve devices at cryogenic temperatures},\ }\href {https://doi.org/10.1063/1.4865464} {\bibfield  {journal} {\bibinfo  {journal} {Journal of Applied Physics}\ }\textbf {\bibinfo {volume} {115}} (\bibinfo {year} {2014})}\BibitemShut {NoStop}%
\bibitem [{\citenamefont {Nguyen}\ \emph {et~al.}(2020)\citenamefont {Nguyen}, \citenamefont {Ribeill}, \citenamefont {Gustafsson}, \citenamefont {Shi}, \citenamefont {Aradhya}, \citenamefont {Wagner}, \citenamefont {Ranzani}, \citenamefont {Zhu}, \citenamefont {Baghdadi}, \citenamefont {Butters} \emph {et~al.}}]{nguyen2020}%
  \BibitemOpen
  \bibfield  {author} {\bibinfo {author} {\bibfnamefont {M.-H.}\ \bibnamefont {Nguyen}}, \bibinfo {author} {\bibfnamefont {G.~J.}\ \bibnamefont {Ribeill}}, \bibinfo {author} {\bibfnamefont {M.~V.}\ \bibnamefont {Gustafsson}}, \bibinfo {author} {\bibfnamefont {S.}~\bibnamefont {Shi}}, \bibinfo {author} {\bibfnamefont {S.~V.}\ \bibnamefont {Aradhya}}, \bibinfo {author} {\bibfnamefont {A.~P.}\ \bibnamefont {Wagner}}, \bibinfo {author} {\bibfnamefont {L.~M.}\ \bibnamefont {Ranzani}}, \bibinfo {author} {\bibfnamefont {L.}~\bibnamefont {Zhu}}, \bibinfo {author} {\bibfnamefont {R.}~\bibnamefont {Baghdadi}}, \bibinfo {author} {\bibfnamefont {B.}~\bibnamefont {Butters}}, \emph {et~al.},\ }\bibfield  {title} {\bibinfo {title} {Cryogenic memory architecture integrating spin hall effect based magnetic memory and superconductive cryotron devices},\ }\href {https://www.nature.com/articles/s41598-019-57137-9} {\bibfield  {journal} {\bibinfo  {journal} {Scientific reports}\ }\textbf {\bibinfo {volume} {10}},\ \bibinfo
  {pages} {248} (\bibinfo {year} {2020})}\BibitemShut {NoStop}%
\bibitem [{\citenamefont {Han}\ \emph {et~al.}(2025)\citenamefont {Han}, \citenamefont {Sun}, \citenamefont {Richstein}, \citenamefont {Grenmyr}, \citenamefont {Bae}, \citenamefont {Allibert}, \citenamefont {Radu}, \citenamefont {Gr\"utzmacher}, \citenamefont {Knoch},\ and\ \citenamefont {Zhao}}]{han2025}%
  \BibitemOpen
  \bibfield  {author} {\bibinfo {author} {\bibfnamefont {Y.}~\bibnamefont {Han}}, \bibinfo {author} {\bibfnamefont {J.}~\bibnamefont {Sun}}, \bibinfo {author} {\bibfnamefont {B.}~\bibnamefont {Richstein}}, \bibinfo {author} {\bibfnamefont {A.}~\bibnamefont {Grenmyr}}, \bibinfo {author} {\bibfnamefont {J.-H.}\ \bibnamefont {Bae}}, \bibinfo {author} {\bibfnamefont {F.}~\bibnamefont {Allibert}}, \bibinfo {author} {\bibfnamefont {I.}~\bibnamefont {Radu}}, \bibinfo {author} {\bibfnamefont {D.}~\bibnamefont {Gr\"utzmacher}}, \bibinfo {author} {\bibfnamefont {J.}~\bibnamefont {Knoch}},\ and\ \bibinfo {author} {\bibfnamefont {Q.-T.}\ \bibnamefont {Zhao}},\ }\bibfield  {title} {\bibinfo {title} {An energy efficient memory cell for quantum and neuromorphic computing at low temperatures},\ }\href {https://pubs.acs.org/doi/pdf/10.1021/acs.nanolett.4c05855} {\bibfield  {journal} {\bibinfo  {journal} {Nano Letters}\ }\textbf {\bibinfo {volume} {25}},\ \bibinfo {pages} {6374} (\bibinfo {year} {2025})}\BibitemShut {NoStop}%
\bibitem [{\citenamefont {Van~Duzer}\ \emph {et~al.}(2012)\citenamefont {Van~Duzer}, \citenamefont {Zheng}, \citenamefont {Whiteley}, \citenamefont {Kim}, \citenamefont {Kim}, \citenamefont {Meng},\ and\ \citenamefont {Ortlepp}}]{duzer2012}%
  \BibitemOpen
  \bibfield  {author} {\bibinfo {author} {\bibfnamefont {T.}~\bibnamefont {Van~Duzer}}, \bibinfo {author} {\bibfnamefont {L.}~\bibnamefont {Zheng}}, \bibinfo {author} {\bibfnamefont {S.~R.}\ \bibnamefont {Whiteley}}, \bibinfo {author} {\bibfnamefont {H.}~\bibnamefont {Kim}}, \bibinfo {author} {\bibfnamefont {J.}~\bibnamefont {Kim}}, \bibinfo {author} {\bibfnamefont {X.}~\bibnamefont {Meng}},\ and\ \bibinfo {author} {\bibfnamefont {T.}~\bibnamefont {Ortlepp}},\ }\bibfield  {title} {\bibinfo {title} {64-kb hybrid josephson-cmos 4 kelvin ram with 400 ps access time and 12 mw read power},\ }\href {https://ieeexplore.ieee.org/abstract/document/6363561} {\bibfield  {journal} {\bibinfo  {journal} {IEEE Transactions on Applied Superconductivity}\ }\textbf {\bibinfo {volume} {23}},\ \bibinfo {pages} {1700504} (\bibinfo {year} {2012})}\BibitemShut {NoStop}%
\bibitem [{\citenamefont {Hironaka}\ \emph {et~al.}(2020)\citenamefont {Hironaka}, \citenamefont {Yamanashi},\ and\ \citenamefont {Yoshikawa}}]{hironaka2020}%
  \BibitemOpen
  \bibfield  {author} {\bibinfo {author} {\bibfnamefont {Y.}~\bibnamefont {Hironaka}}, \bibinfo {author} {\bibfnamefont {Y.}~\bibnamefont {Yamanashi}},\ and\ \bibinfo {author} {\bibfnamefont {N.}~\bibnamefont {Yoshikawa}},\ }\bibfield  {title} {\bibinfo {title} {Demonstration of a single-flux-quantum microprocessor operating with josephson-cmos hybrid memory},\ }\href {https://ieeexplore.ieee.org/abstract/document/9091791} {\bibfield  {journal} {\bibinfo  {journal} {IEEE Transactions on Applied Superconductivity}\ }\textbf {\bibinfo {volume} {30}},\ \bibinfo {pages} {1} (\bibinfo {year} {2020})}\BibitemShut {NoStop}%
\bibitem [{\citenamefont {Volk}\ \emph {et~al.}(2023)\citenamefont {Volk}, \citenamefont {Wynn}, \citenamefont {Golden}, \citenamefont {Sherwood},\ and\ \citenamefont {Tzimpragos}}]{volk2023}%
  \BibitemOpen
  \bibfield  {author} {\bibinfo {author} {\bibfnamefont {J.}~\bibnamefont {Volk}}, \bibinfo {author} {\bibfnamefont {A.}~\bibnamefont {Wynn}}, \bibinfo {author} {\bibfnamefont {E.}~\bibnamefont {Golden}}, \bibinfo {author} {\bibfnamefont {T.}~\bibnamefont {Sherwood}},\ and\ \bibinfo {author} {\bibfnamefont {G.}~\bibnamefont {Tzimpragos}},\ }\bibfield  {title} {\bibinfo {title} {Addressable superconductor integrated circuit memory from delay lines},\ }\href {https://www.nature.com/articles/s41598-023-43205-8} {\bibfield  {journal} {\bibinfo  {journal} {Scientific Reports}\ }\textbf {\bibinfo {volume} {13}},\ \bibinfo {pages} {16639} (\bibinfo {year} {2023})}\BibitemShut {NoStop}%
\bibitem [{\citenamefont {Butters}\ \emph {et~al.}(2021)\citenamefont {Butters}, \citenamefont {Baghdadi}, \citenamefont {Onen}, \citenamefont {Toomey}, \citenamefont {Medeiros},\ and\ \citenamefont {Berggren}}]{butters2021}%
  \BibitemOpen
  \bibfield  {author} {\bibinfo {author} {\bibfnamefont {B.~A.}\ \bibnamefont {Butters}}, \bibinfo {author} {\bibfnamefont {R.}~\bibnamefont {Baghdadi}}, \bibinfo {author} {\bibfnamefont {M.}~\bibnamefont {Onen}}, \bibinfo {author} {\bibfnamefont {E.~A.}\ \bibnamefont {Toomey}}, \bibinfo {author} {\bibfnamefont {O.}~\bibnamefont {Medeiros}},\ and\ \bibinfo {author} {\bibfnamefont {K.~K.}\ \bibnamefont {Berggren}},\ }\bibfield  {title} {\bibinfo {title} {A scalable superconducting nanowire memory cell and preliminary array test},\ }\href {https://iopscience.iop.org/article/10.1088/1361-6668/abd14e/meta} {\bibfield  {journal} {\bibinfo  {journal} {Superconductor Science and Technology}\ }\textbf {\bibinfo {volume} {34}},\ \bibinfo {pages} {035003} (\bibinfo {year} {2021})}\BibitemShut {NoStop}%
\bibitem [{\citenamefont {Buzzi}\ \emph {et~al.}(2023)\citenamefont {Buzzi}, \citenamefont {Castellani}, \citenamefont {Foster}, \citenamefont {Medeiros}, \citenamefont {Colangelo},\ and\ \citenamefont {Berggren}}]{buzzi2023}%
  \BibitemOpen
  \bibfield  {author} {\bibinfo {author} {\bibfnamefont {A.}~\bibnamefont {Buzzi}}, \bibinfo {author} {\bibfnamefont {M.}~\bibnamefont {Castellani}}, \bibinfo {author} {\bibfnamefont {R.~A.}\ \bibnamefont {Foster}}, \bibinfo {author} {\bibfnamefont {O.}~\bibnamefont {Medeiros}}, \bibinfo {author} {\bibfnamefont {M.}~\bibnamefont {Colangelo}},\ and\ \bibinfo {author} {\bibfnamefont {K.~K.}\ \bibnamefont {Berggren}},\ }\bibfield  {title} {\bibinfo {title} {A nanocryotron memory and logic family},\ }\href {https://doi.org/10.1063/5.0144686} {\bibfield  {journal} {\bibinfo  {journal} {Applied Physics Letters}\ }\textbf {\bibinfo {volume} {122}} (\bibinfo {year} {2023})}\BibitemShut {NoStop}%
\bibitem [{\citenamefont {Miyahara}\ \emph {et~al.}(1987)\citenamefont {Miyahara}, \citenamefont {Mukaida}, \citenamefont {Tokumitsu}, \citenamefont {Kubo},\ and\ \citenamefont {Hohkawa}}]{miyahara1987}%
  \BibitemOpen
  \bibfield  {author} {\bibinfo {author} {\bibfnamefont {K.}~\bibnamefont {Miyahara}}, \bibinfo {author} {\bibfnamefont {M.}~\bibnamefont {Mukaida}}, \bibinfo {author} {\bibfnamefont {M.}~\bibnamefont {Tokumitsu}}, \bibinfo {author} {\bibfnamefont {S.}~\bibnamefont {Kubo}},\ and\ \bibinfo {author} {\bibfnamefont {K.}~\bibnamefont {Hohkawa}},\ }\bibfield  {title} {\bibinfo {title} {Abrikosov vortex memory with improved sensitivity and reduced write current levels},\ }\href {https://ieeexplore.ieee.org/abstract/document/1064992} {\bibfield  {journal} {\bibinfo  {journal} {IEEE transactions on magnetics}\ }\textbf {\bibinfo {volume} {23}},\ \bibinfo {pages} {875} (\bibinfo {year} {1987})}\BibitemShut {NoStop}%
\bibitem [{\citenamefont {Golod}\ \emph {et~al.}(2015)\citenamefont {Golod}, \citenamefont {Iovan},\ and\ \citenamefont {Krasnov}}]{golod2015}%
  \BibitemOpen
  \bibfield  {author} {\bibinfo {author} {\bibfnamefont {T.}~\bibnamefont {Golod}}, \bibinfo {author} {\bibfnamefont {A.}~\bibnamefont {Iovan}},\ and\ \bibinfo {author} {\bibfnamefont {V.~M.}\ \bibnamefont {Krasnov}},\ }\bibfield  {title} {\bibinfo {title} {Single abrikosov vortices as quantized information bits},\ }\href {https://www.nature.com/articles/ncomms9628} {\bibfield  {journal} {\bibinfo  {journal} {Nature Communications}\ }\textbf {\bibinfo {volume} {6}},\ \bibinfo {pages} {8628} (\bibinfo {year} {2015})}\BibitemShut {NoStop}%
\bibitem [{\citenamefont {Golod}\ \emph {et~al.}(2023)\citenamefont {Golod}, \citenamefont {Morlet-Decarnin},\ and\ \citenamefont {Krasnov}}]{golod2023}%
  \BibitemOpen
  \bibfield  {author} {\bibinfo {author} {\bibfnamefont {T.}~\bibnamefont {Golod}}, \bibinfo {author} {\bibfnamefont {L.}~\bibnamefont {Morlet-Decarnin}},\ and\ \bibinfo {author} {\bibfnamefont {V.~M.}\ \bibnamefont {Krasnov}},\ }\bibfield  {title} {\bibinfo {title} {Word and bit line operation of a 1$\times$ 1 $\mu$m2 superconducting vortex-based memory},\ }\href {https://www.nature.com/articles/s41467-023-40654-7} {\bibfield  {journal} {\bibinfo  {journal} {Nature Communications}\ }\textbf {\bibinfo {volume} {14}},\ \bibinfo {pages} {4926} (\bibinfo {year} {2023})}\BibitemShut {NoStop}%
\bibitem [{\citenamefont {Nevirkovets}\ and\ \citenamefont {Mukhanov}(2018)}]{nevirkovets2018}%
  \BibitemOpen
  \bibfield  {author} {\bibinfo {author} {\bibfnamefont {I.}~\bibnamefont {Nevirkovets}}\ and\ \bibinfo {author} {\bibfnamefont {O.}~\bibnamefont {Mukhanov}},\ }\bibfield  {title} {\bibinfo {title} {Memory cell for high-density arrays based on a multiterminal superconducting-ferromagnetic device},\ }\href {https://journals.aps.org/prapplied/abstract/10.1103/PhysRevApplied.10.034013} {\bibfield  {journal} {\bibinfo  {journal} {Physical Review Applied}\ }\textbf {\bibinfo {volume} {10}},\ \bibinfo {pages} {034013} (\bibinfo {year} {2018})}\BibitemShut {NoStop}%
\bibitem [{\citenamefont {Nevirkovets}\ and\ \citenamefont {Mukhanov}(2023)}]{nevirkovets2023}%
  \BibitemOpen
  \bibfield  {author} {\bibinfo {author} {\bibfnamefont {I.}~\bibnamefont {Nevirkovets}}\ and\ \bibinfo {author} {\bibfnamefont {O.}~\bibnamefont {Mukhanov}},\ }\bibfield  {title} {\bibinfo {title} {Electrically controlled hybrid superconductor--ferromagnet cell for high density cryogenic memory},\ }\href {https://pubs.aip.org/aip/apl/article/123/7/072601/2906569} {\bibfield  {journal} {\bibinfo  {journal} {Applied Physics Letters}\ }\textbf {\bibinfo {volume} {123}} (\bibinfo {year} {2023})}\BibitemShut {NoStop}%
\bibitem [{\citenamefont {Kryder}\ \emph {et~al.}(2008)\citenamefont {Kryder}, \citenamefont {Gage}, \citenamefont {McDaniel}, \citenamefont {Challener}, \citenamefont {Rottmayer}, \citenamefont {Ju}, \citenamefont {Hsia},\ and\ \citenamefont {Erden}}]{kryder2008}%
  \BibitemOpen
  \bibfield  {author} {\bibinfo {author} {\bibfnamefont {M.~H.}\ \bibnamefont {Kryder}}, \bibinfo {author} {\bibfnamefont {E.~C.}\ \bibnamefont {Gage}}, \bibinfo {author} {\bibfnamefont {T.~W.}\ \bibnamefont {McDaniel}}, \bibinfo {author} {\bibfnamefont {W.~A.}\ \bibnamefont {Challener}}, \bibinfo {author} {\bibfnamefont {R.~E.}\ \bibnamefont {Rottmayer}}, \bibinfo {author} {\bibfnamefont {G.}~\bibnamefont {Ju}}, \bibinfo {author} {\bibfnamefont {Y.-T.}\ \bibnamefont {Hsia}},\ and\ \bibinfo {author} {\bibfnamefont {M.~F.}\ \bibnamefont {Erden}},\ }\bibfield  {title} {\bibinfo {title} {Heat assisted magnetic recording},\ }\href {https://ieeexplore.ieee.org/abstract/document/4694026} {\bibfield  {journal} {\bibinfo  {journal} {Proceedings of the IEEE}\ }\textbf {\bibinfo {volume} {96}},\ \bibinfo {pages} {1810} (\bibinfo {year} {2008})}\BibitemShut {NoStop}%
\bibitem [{\citenamefont {Pagano}\ \emph {et~al.}(2017)\citenamefont {Pagano}, \citenamefont {Martucciello}, \citenamefont {Bobba}, \citenamefont {Carapella}, \citenamefont {Attanasio}, \citenamefont {Cirillo}, \citenamefont {Cristiano}, \citenamefont {Lisitskiy}, \citenamefont {Ejrnaes}, \citenamefont {Pepe} \emph {et~al.}}]{pagano2017}%
  \BibitemOpen
  \bibfield  {author} {\bibinfo {author} {\bibfnamefont {S.}~\bibnamefont {Pagano}}, \bibinfo {author} {\bibfnamefont {N.}~\bibnamefont {Martucciello}}, \bibinfo {author} {\bibfnamefont {F.}~\bibnamefont {Bobba}}, \bibinfo {author} {\bibfnamefont {G.}~\bibnamefont {Carapella}}, \bibinfo {author} {\bibfnamefont {C.}~\bibnamefont {Attanasio}}, \bibinfo {author} {\bibfnamefont {C.}~\bibnamefont {Cirillo}}, \bibinfo {author} {\bibfnamefont {R.}~\bibnamefont {Cristiano}}, \bibinfo {author} {\bibfnamefont {M.}~\bibnamefont {Lisitskiy}}, \bibinfo {author} {\bibfnamefont {M.}~\bibnamefont {Ejrnaes}}, \bibinfo {author} {\bibfnamefont {G.~P.}\ \bibnamefont {Pepe}}, \emph {et~al.},\ }\bibfield  {title} {\bibinfo {title} {Proposal for a nanoscale superconductive memory},\ }\href {https://doi.org/10.1109/TASC.2017.2647903} {\bibfield  {journal} {\bibinfo  {journal} {IEEE Transactions on Applied Superconductivity}\ }\textbf {\bibinfo {volume} {27}},\ \bibinfo {pages} {1} (\bibinfo {year} {2017})}\BibitemShut {NoStop}%
\bibitem [{\citenamefont {De~Gennes}(1966)}]{degennes1966}%
  \BibitemOpen
  \bibfield  {author} {\bibinfo {author} {\bibfnamefont {P.}~\bibnamefont {De~Gennes}},\ }\bibfield  {title} {\bibinfo {title} {Coupling between ferromagnets through a superconducting layer},\ }\href {https://doi.org/10.1016/0031-9163(66)90229-0} {\bibfield  {journal} {\bibinfo  {journal} {Physics Letters}\ }\textbf {\bibinfo {volume} {23}},\ \bibinfo {pages} {10} (\bibinfo {year} {1966})}\BibitemShut {NoStop}%
\bibitem [{\citenamefont {Sarma}(1963)}]{sarma1963}%
  \BibitemOpen
  \bibfield  {author} {\bibinfo {author} {\bibfnamefont {G.}~\bibnamefont {Sarma}},\ }\bibfield  {title} {\bibinfo {title} {On the influence of a uniform exchange field acting on the spins of the conduction electrons in a superconductor},\ }\href {https://doi.org/10.1016/0022-3697(63)90007-6} {\bibfield  {journal} {\bibinfo  {journal} {Journal of Physics and Chemistry of Solids}\ }\textbf {\bibinfo {volume} {24}},\ \bibinfo {pages} {1029} (\bibinfo {year} {1963})}\BibitemShut {NoStop}%
\bibitem [{\citenamefont {Bergeret}\ \emph {et~al.}(2005)\citenamefont {Bergeret}, \citenamefont {Volkov},\ and\ \citenamefont {Efetov}}]{bergeret2005}%
  \BibitemOpen
  \bibfield  {author} {\bibinfo {author} {\bibfnamefont {F.}~\bibnamefont {Bergeret}}, \bibinfo {author} {\bibfnamefont {A.~F.}\ \bibnamefont {Volkov}},\ and\ \bibinfo {author} {\bibfnamefont {K.~B.}\ \bibnamefont {Efetov}},\ }\bibfield  {title} {\bibinfo {title} {Odd triplet superconductivity and related phenomena in superconductor-ferromagnet structures},\ }\href {https://doi.org/10.1103/RevModPhys.77.1321} {\bibfield  {journal} {\bibinfo  {journal} {Reviews of modern physics}\ }\textbf {\bibinfo {volume} {77}},\ \bibinfo {pages} {1321} (\bibinfo {year} {2005})}\BibitemShut {NoStop}%
\bibitem [{\citenamefont {Cai}\ \emph {et~al.}(2023)\citenamefont {Cai}, \citenamefont {{\v{Z}}uti{\'c}},\ and\ \citenamefont {Han}}]{cai2023}%
  \BibitemOpen
  \bibfield  {author} {\bibinfo {author} {\bibfnamefont {R.}~\bibnamefont {Cai}}, \bibinfo {author} {\bibfnamefont {I.}~\bibnamefont {{\v{Z}}uti{\'c}}},\ and\ \bibinfo {author} {\bibfnamefont {W.}~\bibnamefont {Han}},\ }\bibfield  {title} {\bibinfo {title} {Superconductor/ferromagnet heterostructures: a platform for superconducting spintronics and quantum computation},\ }\href {https://onlinelibrary.wiley.com/doi/abs/10.1002/qute.202200080} {\bibfield  {journal} {\bibinfo  {journal} {Advanced Quantum Technologies}\ }\textbf {\bibinfo {volume} {6}},\ \bibinfo {pages} {2200080} (\bibinfo {year} {2023})}\BibitemShut {NoStop}%
\bibitem [{\citenamefont {Hauser}(1969)}]{hauser1969}%
  \BibitemOpen
  \bibfield  {author} {\bibinfo {author} {\bibfnamefont {J.}~\bibnamefont {Hauser}},\ }\bibfield  {title} {\bibinfo {title} {Coupling between ferrimagnetic insulators through a superconducting layer},\ }\href {https://doi.org/10.1103/PhysRevLett.23.374} {\bibfield  {journal} {\bibinfo  {journal} {Physical Review Letters}\ }\textbf {\bibinfo {volume} {23}},\ \bibinfo {pages} {374} (\bibinfo {year} {1969})}\BibitemShut {NoStop}%
\bibitem [{\citenamefont {Li}\ \emph {et~al.}(2013)\citenamefont {Li}, \citenamefont {Roschewsky}, \citenamefont {Assaf}, \citenamefont {Eich}, \citenamefont {Epstein-Martin}, \citenamefont {Heiman}, \citenamefont {M{\"u}nzenberg},\ and\ \citenamefont {Moodera}}]{li2013}%
  \BibitemOpen
  \bibfield  {author} {\bibinfo {author} {\bibfnamefont {B.}~\bibnamefont {Li}}, \bibinfo {author} {\bibfnamefont {N.}~\bibnamefont {Roschewsky}}, \bibinfo {author} {\bibfnamefont {B.~A.}\ \bibnamefont {Assaf}}, \bibinfo {author} {\bibfnamefont {M.}~\bibnamefont {Eich}}, \bibinfo {author} {\bibfnamefont {M.}~\bibnamefont {Epstein-Martin}}, \bibinfo {author} {\bibfnamefont {D.}~\bibnamefont {Heiman}}, \bibinfo {author} {\bibfnamefont {M.}~\bibnamefont {M{\"u}nzenberg}},\ and\ \bibinfo {author} {\bibfnamefont {J.~S.}\ \bibnamefont {Moodera}},\ }\bibfield  {title} {\bibinfo {title} {Superconducting spin switch with infinite magnetoresistance induced by an internal exchange field},\ }\href {https://doi.org/10.1103/PhysRevLett.110.097001} {\bibfield  {journal} {\bibinfo  {journal} {Physical Review Letters}\ }\textbf {\bibinfo {volume} {110}},\ \bibinfo {pages} {097001} (\bibinfo {year} {2013})}\BibitemShut {NoStop}%
\bibitem [{\citenamefont {Zhu}\ \emph {et~al.}(2017)\citenamefont {Zhu}, \citenamefont {Pal}, \citenamefont {Blamire},\ and\ \citenamefont {Barber}}]{zhu2017}%
  \BibitemOpen
  \bibfield  {author} {\bibinfo {author} {\bibfnamefont {Y.}~\bibnamefont {Zhu}}, \bibinfo {author} {\bibfnamefont {A.}~\bibnamefont {Pal}}, \bibinfo {author} {\bibfnamefont {M.~G.}\ \bibnamefont {Blamire}},\ and\ \bibinfo {author} {\bibfnamefont {Z.~H.}\ \bibnamefont {Barber}},\ }\bibfield  {title} {\bibinfo {title} {Superconducting exchange coupling between ferromagnets},\ }\href {https://www.nature.com/articles/nmat4753} {\bibfield  {journal} {\bibinfo  {journal} {Nature Materials}\ }\textbf {\bibinfo {volume} {16}},\ \bibinfo {pages} {195} (\bibinfo {year} {2017})}\BibitemShut {NoStop}%
\bibitem [{\citenamefont {Di~Bernardo}\ \emph {et~al.}(2019)\citenamefont {Di~Bernardo}, \citenamefont {Komori}, \citenamefont {Livanas}, \citenamefont {Divitini}, \citenamefont {Gentile}, \citenamefont {Cuoco},\ and\ \citenamefont {Robinson}}]{dibernardo2019}%
  \BibitemOpen
  \bibfield  {author} {\bibinfo {author} {\bibfnamefont {A.}~\bibnamefont {Di~Bernardo}}, \bibinfo {author} {\bibfnamefont {S.}~\bibnamefont {Komori}}, \bibinfo {author} {\bibfnamefont {G.}~\bibnamefont {Livanas}}, \bibinfo {author} {\bibfnamefont {G.}~\bibnamefont {Divitini}}, \bibinfo {author} {\bibfnamefont {P.}~\bibnamefont {Gentile}}, \bibinfo {author} {\bibfnamefont {M.}~\bibnamefont {Cuoco}},\ and\ \bibinfo {author} {\bibfnamefont {J.~W.}\ \bibnamefont {Robinson}},\ }\bibfield  {title} {\bibinfo {title} {Nodal superconducting exchange coupling},\ }\href {https://www.nature.com/articles/s41563-019-0457-6} {\bibfield  {journal} {\bibinfo  {journal} {Nature materials}\ }\textbf {\bibinfo {volume} {18}},\ \bibinfo {pages} {1194} (\bibinfo {year} {2019})}\BibitemShut {NoStop}%
\bibitem [{\citenamefont {Matsuki}\ \emph {et~al.}(2025)\citenamefont {Matsuki}, \citenamefont {Hijano}, \citenamefont {Mazur}, \citenamefont {Ili{\'c}}, \citenamefont {Wang}, \citenamefont {Alekhina}, \citenamefont {Ohnishi}, \citenamefont {Komori}, \citenamefont {Li}, \citenamefont {Stelmashenko} \emph {et~al.}}]{matsuki2025}%
  \BibitemOpen
  \bibfield  {author} {\bibinfo {author} {\bibfnamefont {H.}~\bibnamefont {Matsuki}}, \bibinfo {author} {\bibfnamefont {A.}~\bibnamefont {Hijano}}, \bibinfo {author} {\bibfnamefont {G.~P.}\ \bibnamefont {Mazur}}, \bibinfo {author} {\bibfnamefont {S.}~\bibnamefont {Ili{\'c}}}, \bibinfo {author} {\bibfnamefont {B.}~\bibnamefont {Wang}}, \bibinfo {author} {\bibfnamefont {I.}~\bibnamefont {Alekhina}}, \bibinfo {author} {\bibfnamefont {K.}~\bibnamefont {Ohnishi}}, \bibinfo {author} {\bibfnamefont {S.}~\bibnamefont {Komori}}, \bibinfo {author} {\bibfnamefont {Y.}~\bibnamefont {Li}}, \bibinfo {author} {\bibfnamefont {N.}~\bibnamefont {Stelmashenko}}, \emph {et~al.},\ }\bibfield  {title} {\bibinfo {title} {Realisation of de gennes’ absolute superconducting switch with a heavy metal interface},\ }\href {https://doi.org/10.1038/s41467-025-61267-2} {\bibfield  {journal} {\bibinfo  {journal} {Nature Communications}\ }\textbf {\bibinfo {volume} {16}},\ \bibinfo {pages} {5674} (\bibinfo {year} {2025})}\BibitemShut
  {NoStop}%
\bibitem [{\citenamefont {Ojaj{\"a}rvi}\ \emph {et~al.}(2022)\citenamefont {Ojaj{\"a}rvi}, \citenamefont {Bergeret}, \citenamefont {Silaev},\ and\ \citenamefont {Heikkil{\"a}}}]{ojajarvi2022}%
  \BibitemOpen
  \bibfield  {author} {\bibinfo {author} {\bibfnamefont {R.}~\bibnamefont {Ojaj{\"a}rvi}}, \bibinfo {author} {\bibfnamefont {F.}~\bibnamefont {Bergeret}}, \bibinfo {author} {\bibfnamefont {M.}~\bibnamefont {Silaev}},\ and\ \bibinfo {author} {\bibfnamefont {T.~T.}\ \bibnamefont {Heikkil{\"a}}},\ }\bibfield  {title} {\bibinfo {title} {Dynamics of two ferromagnetic insulators coupled by superconducting spin current},\ }\href {https://doi.org/10.1103/PhysRevLett.128.167701} {\bibfield  {journal} {\bibinfo  {journal} {Physical Review Letters}\ }\textbf {\bibinfo {volume} {128}},\ \bibinfo {pages} {167701} (\bibinfo {year} {2022})}\BibitemShut {NoStop}%
\bibitem [{\citenamefont {Bhakat}\ \emph {et~al.}(2025)\citenamefont {Bhakat}, \citenamefont {Samanta}, \citenamefont {Mahapatra},\ and\ \citenamefont {Pal}}]{bhakat2025}%
  \BibitemOpen
  \bibfield  {author} {\bibinfo {author} {\bibfnamefont {S.}~\bibnamefont {Bhakat}}, \bibinfo {author} {\bibfnamefont {S.}~\bibnamefont {Samanta}}, \bibinfo {author} {\bibfnamefont {S.}~\bibnamefont {Mahapatra}},\ and\ \bibinfo {author} {\bibfnamefont {A.}~\bibnamefont {Pal}},\ }\bibfield  {title} {\bibinfo {title} {Bistable and absolute switching driven by superconducting exchange coupling},\ }\href {https://www.nature.com/articles/s41467-025-64594-6} {\bibfield  {journal} {\bibinfo  {journal} {Nature Communications}\ }\textbf {\bibinfo {volume} {16}},\ \bibinfo {pages} {9609} (\bibinfo {year} {2025})}\BibitemShut {NoStop}%
\bibitem [{\citenamefont {Ielmini}\ and\ \citenamefont {Wong}(2018)}]{ielmini2018}%
  \BibitemOpen
  \bibfield  {author} {\bibinfo {author} {\bibfnamefont {D.}~\bibnamefont {Ielmini}}\ and\ \bibinfo {author} {\bibfnamefont {H.-S.~P.}\ \bibnamefont {Wong}},\ }\bibfield  {title} {\bibinfo {title} {In-memory computing with resistive switching devices},\ }\href {https://doi.org/10.1038/s41928-018-0092-2} {\bibfield  {journal} {\bibinfo  {journal} {Nature Electronics}\ }\textbf {\bibinfo {volume} {1}},\ \bibinfo {pages} {333} (\bibinfo {year} {2018})}\BibitemShut {NoStop}%
\bibitem [{\citenamefont {Yun}\ \emph {et~al.}(2023)\citenamefont {Yun}, \citenamefont {Son}, \citenamefont {Shin}, \citenamefont {Park}, \citenamefont {Zhang}, \citenamefont {Shin}, \citenamefont {Park},\ and\ \citenamefont {Kim}}]{yun2023}%
  \BibitemOpen
  \bibfield  {author} {\bibinfo {author} {\bibfnamefont {J.}~\bibnamefont {Yun}}, \bibinfo {author} {\bibfnamefont {S.}~\bibnamefont {Son}}, \bibinfo {author} {\bibfnamefont {J.}~\bibnamefont {Shin}}, \bibinfo {author} {\bibfnamefont {G.}~\bibnamefont {Park}}, \bibinfo {author} {\bibfnamefont {K.}~\bibnamefont {Zhang}}, \bibinfo {author} {\bibfnamefont {Y.~J.}\ \bibnamefont {Shin}}, \bibinfo {author} {\bibfnamefont {J.-G.}\ \bibnamefont {Park}},\ and\ \bibinfo {author} {\bibfnamefont {D.}~\bibnamefont {Kim}},\ }\bibfield  {title} {\bibinfo {title} {Magnetic proximity-induced superconducting diode effect and infinite magnetoresistance in a van der waals heterostructure},\ }\href {https://doi.org/10.1103/PhysRevResearch.5.L022064} {\bibfield  {journal} {\bibinfo  {journal} {Physical Review Research}\ }\textbf {\bibinfo {volume} {5}},\ \bibinfo {pages} {L022064} (\bibinfo {year} {2023})}\BibitemShut {NoStop}%
\bibitem [{\citenamefont {Cheng}\ \emph {et~al.}(2025)\citenamefont {Cheng}, \citenamefont {Shu}, \citenamefont {He}, \citenamefont {Dai},\ and\ \citenamefont {Wang}}]{cheng2025}%
  \BibitemOpen
  \bibfield  {author} {\bibinfo {author} {\bibfnamefont {Y.}~\bibnamefont {Cheng}}, \bibinfo {author} {\bibfnamefont {Q.}~\bibnamefont {Shu}}, \bibinfo {author} {\bibfnamefont {H.}~\bibnamefont {He}}, \bibinfo {author} {\bibfnamefont {B.}~\bibnamefont {Dai}},\ and\ \bibinfo {author} {\bibfnamefont {K.~L.}\ \bibnamefont {Wang}},\ }\bibfield  {title} {\bibinfo {title} {Current-driven magnetization switching for superconducting diode memory},\ }\href {https://advanced.onlinelibrary.wiley.com/doi/full/10.1002/adma.202415480} {\bibfield  {journal} {\bibinfo  {journal} {Advanced Materials}\ }\textbf {\bibinfo {volume} {37}},\ \bibinfo {pages} {2415480} (\bibinfo {year} {2025})}\BibitemShut {NoStop}%
\bibitem [{\citenamefont {Mukhanov}\ \emph {et~al.}(1987)\citenamefont {Mukhanov}, \citenamefont {Semenov},\ and\ \citenamefont {Likharev}}]{mukhanov1987}%
  \BibitemOpen
  \bibfield  {author} {\bibinfo {author} {\bibfnamefont {O.}~\bibnamefont {Mukhanov}}, \bibinfo {author} {\bibfnamefont {V.}~\bibnamefont {Semenov}},\ and\ \bibinfo {author} {\bibfnamefont {K.}~\bibnamefont {Likharev}},\ }\bibfield  {title} {\bibinfo {title} {Ultimate performance of the rsfq logic circuits},\ }\href {https://ieeexplore.ieee.org/abstract/document/1064951} {\bibfield  {journal} {\bibinfo  {journal} {IEEE Transactions on Magnetics}\ }\textbf {\bibinfo {volume} {23}},\ \bibinfo {pages} {759} (\bibinfo {year} {1987})}\BibitemShut {NoStop}%
\bibitem [{\citenamefont {Hou}\ \emph {et~al.}(2023)\citenamefont {Hou}, \citenamefont {Nichele}, \citenamefont {Chi}, \citenamefont {Lodesani}, \citenamefont {Wu}, \citenamefont {Ritter}, \citenamefont {Haxell}, \citenamefont {Davydova}, \citenamefont {Ili{\'c}}, \citenamefont {Glezakou-Elbert} \emph {et~al.}}]{hou2023}%
  \BibitemOpen
  \bibfield  {author} {\bibinfo {author} {\bibfnamefont {Y.}~\bibnamefont {Hou}}, \bibinfo {author} {\bibfnamefont {F.}~\bibnamefont {Nichele}}, \bibinfo {author} {\bibfnamefont {H.}~\bibnamefont {Chi}}, \bibinfo {author} {\bibfnamefont {A.}~\bibnamefont {Lodesani}}, \bibinfo {author} {\bibfnamefont {Y.}~\bibnamefont {Wu}}, \bibinfo {author} {\bibfnamefont {M.~F.}\ \bibnamefont {Ritter}}, \bibinfo {author} {\bibfnamefont {D.~Z.}\ \bibnamefont {Haxell}}, \bibinfo {author} {\bibfnamefont {M.}~\bibnamefont {Davydova}}, \bibinfo {author} {\bibfnamefont {S.}~\bibnamefont {Ili{\'c}}}, \bibinfo {author} {\bibfnamefont {O.}~\bibnamefont {Glezakou-Elbert}}, \emph {et~al.},\ }\bibfield  {title} {\bibinfo {title} {Ubiquitous superconducting diode effect in superconductor thin films},\ }\href {https://journals.aps.org/prl/abstract/10.1103/PhysRevLett.131.027001} {\bibfield  {journal} {\bibinfo  {journal} {Physical Review Letters}\ }\textbf {\bibinfo {volume} {131}},\ \bibinfo {pages} {027001} (\bibinfo {year}
  {2023})}\BibitemShut {NoStop}%
\bibitem [{\citenamefont {Suri}\ \emph {et~al.}(2022)\citenamefont {Suri}, \citenamefont {Kamra}, \citenamefont {Meier}, \citenamefont {Kronseder}, \citenamefont {Belzig}, \citenamefont {Back},\ and\ \citenamefont {Strunk}}]{suri2022}%
  \BibitemOpen
  \bibfield  {author} {\bibinfo {author} {\bibfnamefont {D.}~\bibnamefont {Suri}}, \bibinfo {author} {\bibfnamefont {A.}~\bibnamefont {Kamra}}, \bibinfo {author} {\bibfnamefont {T.~N.}\ \bibnamefont {Meier}}, \bibinfo {author} {\bibfnamefont {M.}~\bibnamefont {Kronseder}}, \bibinfo {author} {\bibfnamefont {W.}~\bibnamefont {Belzig}}, \bibinfo {author} {\bibfnamefont {C.~H.}\ \bibnamefont {Back}},\ and\ \bibinfo {author} {\bibfnamefont {C.}~\bibnamefont {Strunk}},\ }\bibfield  {title} {\bibinfo {title} {Non-reciprocity of vortex-limited critical current in conventional superconducting micro-bridges},\ }\href {https://pubs.aip.org/aip/apl/article/121/10/102601/2834233/Non-reciprocity-of-vortex-limited-critical-current} {\bibfield  {journal} {\bibinfo  {journal} {Applied Physics Letters}\ }\textbf {\bibinfo {volume} {121}} (\bibinfo {year} {2022})}\BibitemShut {NoStop}%
\bibitem [{\citenamefont {Ingla-Ayn{\'e}s}\ \emph {et~al.}(2025)\citenamefont {Ingla-Ayn{\'e}s}, \citenamefont {Hou}, \citenamefont {Wang}, \citenamefont {Chu}, \citenamefont {Mukhanov}, \citenamefont {Wei},\ and\ \citenamefont {Moodera}}]{ingla2025}%
  \BibitemOpen
  \bibfield  {author} {\bibinfo {author} {\bibfnamefont {J.}~\bibnamefont {Ingla-Ayn{\'e}s}}, \bibinfo {author} {\bibfnamefont {Y.}~\bibnamefont {Hou}}, \bibinfo {author} {\bibfnamefont {S.}~\bibnamefont {Wang}}, \bibinfo {author} {\bibfnamefont {E.-D.}\ \bibnamefont {Chu}}, \bibinfo {author} {\bibfnamefont {O.~A.}\ \bibnamefont {Mukhanov}}, \bibinfo {author} {\bibfnamefont {P.}~\bibnamefont {Wei}},\ and\ \bibinfo {author} {\bibfnamefont {J.~S.}\ \bibnamefont {Moodera}},\ }\bibfield  {title} {\bibinfo {title} {Efficient superconducting diodes and rectifiers for quantum circuitry},\ }\href {https://doi.org/10.1038/s41928-025-01375-5} {\bibfield  {journal} {\bibinfo  {journal} {Nature Electronics}\ }\textbf {\bibinfo {volume} {8}},\ \bibinfo {pages} {411} (\bibinfo {year} {2025})}\BibitemShut {NoStop}%
\bibitem [{\citenamefont {Nadeem}\ \emph {et~al.}(2023)\citenamefont {Nadeem}, \citenamefont {Fuhrer},\ and\ \citenamefont {Wang}}]{nadeem2023}%
  \BibitemOpen
  \bibfield  {author} {\bibinfo {author} {\bibfnamefont {M.}~\bibnamefont {Nadeem}}, \bibinfo {author} {\bibfnamefont {M.~S.}\ \bibnamefont {Fuhrer}},\ and\ \bibinfo {author} {\bibfnamefont {X.}~\bibnamefont {Wang}},\ }\bibfield  {title} {\bibinfo {title} {The superconducting diode effect},\ }\href {https://www.nature.com/articles/s42254-023-00632-w} {\bibfield  {journal} {\bibinfo  {journal} {Nature Reviews Physics}\ }\textbf {\bibinfo {volume} {5}},\ \bibinfo {pages} {558} (\bibinfo {year} {2023})}\BibitemShut {NoStop}%
\bibitem [{\citenamefont {Castellani}\ \emph {et~al.}(2025)\citenamefont {Castellani}, \citenamefont {Medeiros}, \citenamefont {Buzzi}, \citenamefont {Foster}, \citenamefont {Colangelo},\ and\ \citenamefont {Berggren}}]{castellani2025}%
  \BibitemOpen
  \bibfield  {author} {\bibinfo {author} {\bibfnamefont {M.}~\bibnamefont {Castellani}}, \bibinfo {author} {\bibfnamefont {O.}~\bibnamefont {Medeiros}}, \bibinfo {author} {\bibfnamefont {A.}~\bibnamefont {Buzzi}}, \bibinfo {author} {\bibfnamefont {R.~A.}\ \bibnamefont {Foster}}, \bibinfo {author} {\bibfnamefont {M.}~\bibnamefont {Colangelo}},\ and\ \bibinfo {author} {\bibfnamefont {K.~K.}\ \bibnamefont {Berggren}},\ }\bibfield  {title} {\bibinfo {title} {A superconducting full-wave bridge rectifier},\ }\href {https://doi.org/10.1038/s41928-025-01376-4} {\bibfield  {journal} {\bibinfo  {journal} {Nature Electronics}\ }\textbf {\bibinfo {volume} {8}},\ \bibinfo {pages} {417–425} (\bibinfo {year} {2025})}\BibitemShut {NoStop}%
\bibitem [{\citenamefont {Li}\ \emph {et~al.}(2025)\citenamefont {Li}, \citenamefont {Zhang}, \citenamefont {Wang}, \citenamefont {He}, \citenamefont {Lyu}, \citenamefont {Wang}, \citenamefont {Dong}, \citenamefont {Zhu}, \citenamefont {Matsuki}, \citenamefont {Zhu} \emph {et~al.}}]{li2025}%
  \BibitemOpen
  \bibfield  {author} {\bibinfo {author} {\bibfnamefont {J.}~\bibnamefont {Li}}, \bibinfo {author} {\bibfnamefont {Z.}~\bibnamefont {Zhang}}, \bibinfo {author} {\bibfnamefont {S.}~\bibnamefont {Wang}}, \bibinfo {author} {\bibfnamefont {Y.}~\bibnamefont {He}}, \bibinfo {author} {\bibfnamefont {H.}~\bibnamefont {Lyu}}, \bibinfo {author} {\bibfnamefont {Q.}~\bibnamefont {Wang}}, \bibinfo {author} {\bibfnamefont {B.}~\bibnamefont {Dong}}, \bibinfo {author} {\bibfnamefont {D.}~\bibnamefont {Zhu}}, \bibinfo {author} {\bibfnamefont {H.}~\bibnamefont {Matsuki}}, \bibinfo {author} {\bibfnamefont {D.}~\bibnamefont {Zhu}}, \emph {et~al.},\ }\bibfield  {title} {\bibinfo {title} {Field-free superconducting diode enabled by geometric asymmetry and perpendicular magnetization},\ }\href {https://arxiv.org/abs/2506.17651} {\bibfield  {journal} {\bibinfo  {journal} {arXiv preprint arXiv:2506.17651}\ } (\bibinfo {year} {2025})}\BibitemShut {NoStop}%
\bibitem [{\citenamefont {Golod}\ and\ \citenamefont {Krasnov}(2022)}]{golod2022}%
  \BibitemOpen
  \bibfield  {author} {\bibinfo {author} {\bibfnamefont {T.}~\bibnamefont {Golod}}\ and\ \bibinfo {author} {\bibfnamefont {V.~M.}\ \bibnamefont {Krasnov}},\ }\bibfield  {title} {\bibinfo {title} {Demonstration of a superconducting diode-with-memory, operational at zero magnetic field with switchable nonreciprocity},\ }\href {https://www.nature.com/articles/s41467-022-31256-w} {\bibfield  {journal} {\bibinfo  {journal} {Nature Communications}\ }\textbf {\bibinfo {volume} {13}},\ \bibinfo {pages} {3658} (\bibinfo {year} {2022})}\BibitemShut {NoStop}%
\bibitem [{\citenamefont {Vodolazov}\ and\ \citenamefont {Peeters}(2005)}]{2vodolazov2005}%
  \BibitemOpen
  \bibfield  {author} {\bibinfo {author} {\bibfnamefont {D.~Y.}\ \bibnamefont {Vodolazov}}\ and\ \bibinfo {author} {\bibfnamefont {F.}~\bibnamefont {Peeters}},\ }\bibfield  {title} {\bibinfo {title} {Superconducting rectifier based on the asymmetric surface barrier effect},\ }\href {https://journals.aps.org/prb/abstract/10.1103/PhysRevB.72.172508} {\bibfield  {journal} {\bibinfo  {journal} {Physical Review B}\ }\textbf {\bibinfo {volume} {72}},\ \bibinfo {pages} {172508} (\bibinfo {year} {2005})}\BibitemShut {NoStop}%
\bibitem [{\citenamefont {Manna}\ \emph {et~al.}(2020)\citenamefont {Manna}, \citenamefont {Wei}, \citenamefont {Xie}, \citenamefont {Law}, \citenamefont {Lee},\ and\ \citenamefont {Moodera}}]{manna2020}%
  \BibitemOpen
  \bibfield  {author} {\bibinfo {author} {\bibfnamefont {S.}~\bibnamefont {Manna}}, \bibinfo {author} {\bibfnamefont {P.}~\bibnamefont {Wei}}, \bibinfo {author} {\bibfnamefont {Y.}~\bibnamefont {Xie}}, \bibinfo {author} {\bibfnamefont {K.~T.}\ \bibnamefont {Law}}, \bibinfo {author} {\bibfnamefont {P.~A.}\ \bibnamefont {Lee}},\ and\ \bibinfo {author} {\bibfnamefont {J.~S.}\ \bibnamefont {Moodera}},\ }\bibfield  {title} {\bibinfo {title} {Signature of a pair of majorana zero modes in superconducting gold surface states},\ }\href {https://www.pnas.org/doi/abs/10.1073/pnas.1919753117} {\bibfield  {journal} {\bibinfo  {journal} {Proceedings of the National Academy of Sciences}\ }\textbf {\bibinfo {volume} {117}},\ \bibinfo {pages} {8775} (\bibinfo {year} {2020})}\BibitemShut {NoStop}%
\end{thebibliography}%

\end{document}